\documentclass[aps,prd,
showpage,nobibnotes,groupaddress,amssymb,amsmath,nofootinbib,floatfix]{revtex4-2}

\usepackage{graphicx,amsfonts,amssymb}
\usepackage{amsmath}
\usepackage{epsfig}
\usepackage{dcolumn}
\usepackage{bm}
\usepackage{siunitx}
\usepackage[hidelinks]{hyperref}
\usepackage{float}
\usepackage{array,booktabs}
\usepackage{mathrsfs}
\usepackage{adjustbox}
\usepackage{hyperref}
\usepackage{float}
\usepackage{makecell}
\usepackage{dsfont}
\usepackage{afterpage}

\allowdisplaybreaks

\begin{document}

\title{On the 3-3-1 Landau pole}


\author{Mario W. Barela}
\email{mario.barela@unesp.br}
\affiliation{
Instituto de F\'\i sica Te\'orica, Universidade Estadual Paulista, \\
R. Dr. Bento Teobaldo Ferraz 271, Barra Funda\\ S\~ao Paulo - SP, 01140-070,
Brazil}

\begin{abstract}%
The Minimal 3-3-1 Model (m331) offers a compelling extension of the Standard Model, with a rich particle content that allows for exciting phenomenological possibilities. The parameter space of the theory, however, remains relatively unconstrained. Regarding its effective range of validity, it is understood in the literature that the $U(1)_X$ coupling diverges at around $\SI{4}{TeV}$, which threatens the character of the theory as fundamental. In this work, we rederive the running of the electroweak gauge couplings through a more precise, effective approach, in which the complete parametrization of the full 3-3-1 symmetry is considered at all energies. We show that this more rigorous description guarantees a larger perturbative range for the m331, with a most conservative upper limit going up to around $\SI{8.5}{TeV}$. We also reconsider the usefulness of a relation which eliminates a free parameter of the model in favour of known quantities, regularly invoked within the m331 context. 
\end{abstract}

\maketitle

\section{Introduction}
\label{sec:intro}

Although a remarkably successful theory of the elementary particles and fundamental forces, the Standard Model (SM) is understood to be an incomplete description of nature. The absence of neutrino masses; The inability of accounting for cosmological expectations such as dark matter or baryogenesis; And the discomfort that emerges from the hierarchy or strong $\mathit{CP}$ problem are some of the reasons why new physics is needed.

Among the most notable candidates to replace or improve the SM appeared the classes of Supersymmetric~\cite{Martin:1997ns} and Grand Unified Theories~\cite{Buras:1977yy,Langacker:1980js}, soon to be frustrated or pushed to higher energy regimes by experimental data~\cite{ATLAS:2019lng,CMS:2019zmd}. A contrasting type of models, built to have lower characteristic energies, is marked by having non-simple gauge groups. Known instances of this are the SM plus a $Z^\prime$~\cite{Feldman:2007wj,Kors:2004dx,He:1991qd}, an $SU(4)_L \times U(1)_X$  electroweak model~\cite{Pisano:1994tf}, left-right symmetric electroweak sectors \cite{Senjanovic:1975rk,Mohapatra:1974hk,Deshpande:1990ip}, and the theory in which we are interested, the 3-3-1 Model~\cite{Pisano:1992bxx,Frampton:1992wt}.

Originally proposed as a natural cure to the non-unitarity of right-handed currents coupled to an exotic $W^\prime$~\cite{Pisano:1992bxx}, several versions of 3-3-1 models have been explored since. Motivated examples are one with right-handed neutrinos~\cite{PhysRevD.47.2918,PhysRevD.50.R34,PhysRevD.53.437}, one featuring heavy exotic charged leptons~\cite{PhysRevD.48.2353} and the Minimal 3-3-1 Model (m331), the closest to the first constructions~\cite{Pisano:1992bxx,Frampton:1992wt}. Like other versions, the m331 presents possibilities of solution to several issues of the SM, such as, with small modifications to the representation content, being capable of accommodating Dark Model candidates~\cite{Dong:2014esa,Dong:2015rka,Huong:2019vej}; Allowing for coupling constant unification even without supersymmetrization~\cite{Deppisch:2016jzl,Diaz:2005bw}; And providing an explanation for the arbitrariness in the number of families. In particular, the m331 stands out by containing a doubly-charged vector bilepton $U^{\pm \pm}$, rare feature of BSM models that exists, besides the m331, only in a $SU(15)$ GUT~\cite{PhysRevLett.64.619,Pal:1991nfm}. This was another motivation of the original proponents of the model~\cite{Frampton:1992wt}.

The name of the theory derives from its gauge group,

\begin{equation}
SU(3)_c \times SU(3)_L \times U(1)_X,
\end{equation}
whose $U(1)_X$ coupling, like the coupling of the abelian $U(1)_Y$ of the SM, grows with energy through the Renormalization Group Equation (RGE). However, unlike $g_Y$, which poses no threat as it diverges at 'unreachable' scales, $g_X$ is understood to present arbitrary growth (sometimes called Landau Pole) at around $\SI{4}{TeV}$~\cite{Dias:2004dc,Martinez:2006gb,Santos:2017jbv,Doff:2023bgy}. If unchanged by additional theoretical mechanisms such as emerging states in the theory, this fact severely limits the usefulness of the model, as it implies that perturbativity is lost at nearly experimentable low TeV scales. 

The Spontaneous Symmetry Breaking (SSB) of the m331 decomposes into two stages: the acquisition of a vacuum expectation value (VEV) by the $\chi_0$ scalar triggers  

\begin{equation}
SU(3)_c \times SU(3)_L \times U(1)_X \to SU(3)_c \times SU(2)_L \times U(1)_{Y},
\end{equation} 
and the condensation of every additional neutral scalar in the model causes the SM breaking

\begin{equation}
SU(3)_c \times SU(2)_L \times U(1)_Y \to SU(3)_c \times U(1)_{\text{EM}}.
\end{equation}
This SSB pattern is naturally employed in the investigation of the runnings of the various quantities of the model. In this work, we reexamine this practice and the matter of the m331 Landau Pole, rederiving it in this usual fashion in detail, and stressing the way in which it is dependent on the placement of the breaking scales. More importantly, we call attention to the fact that this approach is merely a useful approximation, specially appropriate at low energies. We effect detailed calculations within the more accurate framework, considering the 3-3-1 symmetry from the beggining and integrating out the exact heavy eigenstates, obtaining that, dependending on the scalar VEVs, the perturbativity range of the theory is at least two times larger than the traditional expectation, but may be arbitrarily extended further with increasing $v_\chi$, the VEV of the $\chi$ triplet. We focus solely on the minimal version of the model (m331), and similar analysis for other versions are left for subsequent works.

The paper is organized as follows. In Section~\ref{sec:matching} we present a brief introduction to symmetry matching conditions and discuss some probable inaccuracies; In Section~\ref{sec:331} we perform a  review of the necessary aspects of the m331; In Section~\ref{sec:ext} we review the usual method of running the constants, approximating the 3-3-1 by an intermediate SM symmetry; In Section~\ref{sec:alt} we describe the full method of thoroughly calculating the evolution of the couplings within the m331, and present the main results of this work; And our conclusions and perspective are discussed in Section~\ref{sec:con}.

\section{Symmetry Matching Conditions and Symmetric RGE}
\label{sec:matching}

Matching theories across thresholds at which the symmetry is altered is a topic tied to some confusion, which is why we succinctly discuss it. Consider the breaking pattern $G_1 \times G_2 \times \cdots \times G_n \to G$, where all factors are simple, with respective coupling constants $g_1,g_2,\cdots g_n$ and $g$. In this case, at the breaking scale, the matching conditions for $g$ usually fall into one of the following two categories \cite{Georgi:1977wk}:

\begin{enumerate}
\item $G \subset G_1$

If the embedding is such that the lower group is entirely contained within a simple higher group factor, the condition becomes simply

\begin{equation}
g = g_1.
\end{equation}

\item $G \subset G_1 \times G_2 \times \cdots G_n$

Suppose, more generally, that $G$ is contained within the non-simple product and specialize to the case $G = U(1)$. The embedding may be parametrized as\footnote{Clearly, the product of simple groups in which $G$ is embedded could be a proper subset of the complete symmetry above the threshold, in which case $p_i = 0$ for some $i$.}

\begin{equation}
Z = \sum_{i=1}^n p_i T_i,
\end{equation}
where $Z$ is the $U(1)$ generator and $T_i$ collectively denote every generator of $G_1 \times G_2 \times \cdots G_n$. In this case, the matching condition reads\footnote{Notice that, in the sum above, every distinct $i$-term which corresponds to generators of a same group exhibits the same coupling factor, \textit{i.e.}, $g_i = g_j$ if $T_i,T_j$ belong to the algebra of the same simple group.}

\begin{equation} \label{eq:matchEq}
\frac{1}{g^2} = \sum_{i=1}^n \frac{p_i^2}{g_i^2}.
\end{equation}
\end{enumerate}

These conditions are not analogous to the finding of Wilson coefficients in the matching of effective theories at mass thresholds. They are, rather, simply a consequence of requiring that the theory can be described by Lagrangians with the expected symmetry and corresponding couplings at each energy range, and that at the breaking scale the theories coincide, as required by continuity. A simple example is the SM relation

\begin{equation} \label{eq:SMMatching}
\frac{1}{e^2} = \frac{1}{g_{2L}^2} + \frac{1}{g_{Y}^2},
\end{equation}
 which is a consequence of the definition of electric charge and that may, at the level of the Lagrangian, be read from 

\begin{equation}
Q = T_3 + Y.
\end{equation}

Once the different couplings, defined to exist at distinct energy ranges, are correctly matched at the symmetry transition scales, they may be evolved through the RGE between thresholds or towards infinity according to

\begin{equation}
\beta(g) = \mu \frac{dg}{d\mu}.
\end{equation}
The $\beta$-function {\it in a fully gauge-symmetric theory} may be written as

\begin{equation} \label{eq:betaeq}
\beta(g) = -\frac{g^3}{(4 \pi)^2}b_1,
\end{equation}
where the $\beta$-function coefficient at one loop may be readily found from its related group theoretical quantities \cite{Roy:2019jqs}

\begin{equation} \label{eq:betacoeff}
b_1 = \frac{11}{3}C_2(\text{Gauge}) - \frac{4}{3}\kappa S_2 (\text{Fermion}) - \frac{1}{6} \eta S_2(\text{Scalar}),
\end{equation} 
where $C_2(R)$ and $S_2(R)$ are, respectively, the Casimir and Dynkin Index invariants of the representation $R$, and $\kappa=1/2(1)$ for Weyl(Dirac) components and $\eta = 1(2)$ for real(complex) scalars.

\section{331 Review}
\label{sec:331}

For our purposes, a sufficient review of the m331 is required. The gauge group is $SU(3)_c \times SU(3)_L \times U(1)_X$, with the electric charge operator in general 3-3-1 models given as usual by a combination of the diagonal generators 

\begin{equation} \label{eq:charge}
Q = T_3 - \beta T_8 + X,
\end{equation}
ensuring that the directions of the defined multiplets are charge eigenstates. The $SU(3)_L$ generators are $T^a = \frac{\lambda^a}{2}$, where $\lambda^a$ are the Gell-Mann matrices. The representation content of the m331 is accommodated by $\beta = \sqrt{3}$, and the `minimal' in m331 stems from the fact that this is the version in which the leptonic particle content is identical to that of the SM, without new particles.  

The representation content of the leptonic sector is formed by

\begin{equation}
L_{\ell}\equiv \begin{pmatrix}
\nu_\ell \\
\ell \\
\ell^c
\end{pmatrix}_{\hspace{-1.3mm}L} \thicksim (\mathbf{1},\mathbf{3},0), \;\;\;\; \ell=e,\mu,\tau.
\end{equation}
In this model, every degree of freedom of any leptonic generation is contained within a single triplet. 

The quark sector is composed by
\begin{gather}
Q_{iL}\equiv \begin{pmatrix}
d_i \\
-u_i \\
j_i
\end{pmatrix}_{\hspace{-1.3mm}L} \hspace{-1.4mm} \thicksim (\mathbf{3},\mathbf{\overline{3}},-1/3), \;\;\; Q_{3L}\equiv\begin{pmatrix}
u_3 \\
d_3 \\
J
\end{pmatrix}_{\hspace{-1.3mm}L} \hspace{-1.4mm} \thicksim (\mathbf{3},\mathbf{3},2/3)  \nonumber \\ 
u_{\alpha R} \thicksim \left(\mathbf{3},\mathbf{1},2/3\right), \;\;\;\; d_{\alpha R} \thicksim \left(\mathbf{3},\mathbf{1},-1/3\right) \\
J_{R} \thicksim \left(\mathbf{3},\mathbf{1},5/3\right), \;\;\;\; j_{i R} \thicksim \left(\mathbf{3},\mathbf{1},-4/3\right) \nonumber \\
i=1,2; \;\;\;\; \alpha=1,2,3. \nonumber
\end{gather}
Notice that $J^{+5/3}$ and $j_{1,2}^{-4/3}$ are exotic quarks.

The scalar sector is comprised, in the usual minimal model, by three triplets

\begin{equation}
\begin{split}
\eta\equiv \begin{pmatrix}
\eta^0 \\
\eta_1^- \\
\eta_2^+
\end{pmatrix}_{\hspace{-1.3mm}L}& \hspace{-1.4mm} \thicksim (\mathbf{1},\mathbf{3},0), \;\;\;\;\;\; \rho\equiv\begin{pmatrix}
\rho^+ \\
\rho^0 \\
\rho^{++}
\end{pmatrix}_{\hspace{-1.3mm}L} \hspace{-1.4mm} \thicksim (\mathbf{1},\mathbf{3},1) \\ 
 \chi&\equiv\begin{pmatrix}
\chi^- \\
\chi^{--} \\
\chi^0
\end{pmatrix}_{\hspace{-1.3mm}L} \hspace{-1.4mm} \thicksim (\mathbf{1},\mathbf{3},-1),
\end{split}
\end{equation}
and a sextet

\begin{equation}
S \equiv \begin{pmatrix}
\sigma_1^0 & \frac{h_2^+}{\sqrt{2}} & \frac{h_1^-}{\sqrt{2}} \\
\frac{h_2^+}{\sqrt{2}} & H_1^{++} & \frac{\sigma_2^0}{\sqrt{2}} \\
\frac{h_1^-}{\sqrt{2}} & \frac{\sigma_2^0}{\sqrt{2}} & H_2^{--}
\end{pmatrix} \thicksim (\mathbf{1},\mathbf{\overline{6}},0).
\end{equation}
This composition for the scalar sector is deemed minimal for being the smallest construction sufficient to fit 
fermion masses, which can be seen in the following way. The $\chi$ is immediately indispensable as it is necessary for the SSB, and at least one among the $X=0$ multiplets, $S$ and $\eta$, is needed to construct a leptonic Yukawa invariant. With both present, the Yukawa terms available are

\begin{equation}\label{eq:YukawaLeptons}
\mathcal{L}_\ell^\text{Y}= \frac{1}{2}G_{ab}^\eta \overline{(L_{ai})^c}L_{bj} \epsilon^{ijk} \eta_k + \frac{1}{2}G_{ab}^S \overline{(L_{ai})^c}L_{bj} S^{ij}+ \text{H.C.},
\end{equation}
with $G^\eta$ an anti-symmetric and $G^S$ a symmetric matrix. If only the $\eta$ was included, the lepton mass matrix would be anti-symmetric, with a spectrum (after a chiral rotation) of the form $\{0,m,m\}$, which demonstrates that the sextet is required. In the other hand, if the $\eta$ was omitted, the lepton and the neutrino mass matrices generated by the equation above would be proportional $M_\ell \propto M_\nu$, and thus diagonalized by the same transformation. In turn, this results in a PMNS matrix equal to the identity, also phenomenologically unsuitable. Finally, the $\rho$ triplet is necessary to fit the known quark masses.

The projection of the complete matter content of the m331 onto the SM appears in Table~\ref{tab:SMProjection}. Note that the entire exotic sector appears as closed multiplets. 

\begin{table}[t!]
\caption{Representation content of the m331 projected onto the SM symmetry.}
\begin{center}
\begin{tabular}{cc}
\toprule
Source multiplet	 & Projected multiplets
 \\ \midrule
$L_\ell$ & $L^{(2)}_{\ell}\equiv
\begin{pmatrix}
\nu_\ell \\
\ell 
\end{pmatrix}_{\hspace{-1.3mm}L} \hspace{-1.4mm} \thicksim (\mathbf{1},\mathbf{2},-1/2) \;\;\;\; \ell_R \thicksim \left(\mathbf{1},\mathbf{1},-1\right)$ \\ \midrule
$Q_{iL}$ &  \makecell{$Q^{(2)}_{iL}\equiv \begin{pmatrix}
d_i \\
u_i 
\end{pmatrix}_{\hspace{-1.3mm}L} \hspace{-1.4mm} \thicksim (\mathbf{3},\mathbf{\overline{2}},1/6) \;\;\;\; u_{i R} \thicksim \left(\mathbf{3},\mathbf{1},2/3\right) \;\;\;\; d_{i R} \thicksim \left(\mathbf{3},\mathbf{1},-1/3\right)$ \\
$j_{i L} \thicksim \left(\mathbf{3},\mathbf{1},-4/3\right) \;\;\;\; j_{i R} \thicksim \left(\mathbf{3},\mathbf{1},-4/3\right)$} \\ \midrule
$Q_{3L}$ &  \makecell{$Q^{(2)}_{3L}\equiv \begin{pmatrix}
u_3 \\
d_3 
\end{pmatrix}_{\hspace{-1.3mm}L} \hspace{-1.4mm} \thicksim (\mathbf{3},\mathbf{2},1/6) \;\;\;\; u_{3 R} \thicksim \left(\mathbf{3},\mathbf{1},2/3\right) \;\;\;\; d_{3 R} \thicksim \left(\mathbf{3},\mathbf{1},-1/3\right)$ \\ 
$J_{ L} \thicksim \left(\mathbf{3},\mathbf{1},5/3\right) \;\;\;\; J_{R} \thicksim \left(\mathbf{3},\mathbf{1},5/3\right)$} \\ \midrule
$\eta$   &  $\phi_\eta \equiv \begin{pmatrix}
\eta^0 \\
\eta_1^- 
\end{pmatrix}_{\hspace{-1.3mm}L} \hspace{-1.4mm}\thicksim (\mathbf{1},\mathbf{2},-1/2) \;\;\;\; \eta^{+}_2 \thicksim (\mathbf{1},\mathbf{1},1)$ \\ \midrule
$\rho$   &   $\phi_\rho\equiv \begin{pmatrix}
\rho^+ \\
\rho^0
\end{pmatrix}_{\hspace{-1.3mm}L} \hspace{-1.4mm} \thicksim (\mathbf{1},\mathbf{2},1/2) \;\;\;\; \rho^{++} \thicksim (\mathbf{1},\mathbf{1},2)$ \\ \midrule
$\chi$    &    $\phi_\chi\equiv\begin{pmatrix}
\chi^- \\
\chi^{--}
\end{pmatrix}_{\hspace{-1.3mm}L} \hspace{-1.4mm} \thicksim (\mathbf{1},\mathbf{2},-3/2) \;\;\;\; \chi^{0} \thicksim (\mathbf{1},\mathbf{1},0)$ \\ \midrule
$S$   &   \makecell{$\phi_S\equiv \begin{pmatrix}
h_1^- \\
\sigma_2^{0}
\end{pmatrix}_{\hspace{-1.3mm}L} \hspace{-1.4mm} \thicksim (\mathbf{1},\mathbf{\overline{2}},-1/2) \;\;\;\; H^{--} \thicksim (\mathbf{1},\mathbf{1},-2)$ \\
$\Phi_S \equiv \begin{pmatrix}
\sigma_1^0 & \frac{h_2^+}{\sqrt{2}} \\
\frac{h_2^+}{\sqrt{2}} & H_1^{++}
\end{pmatrix} \thicksim (\mathbf{1},\mathbf{\overline{3}},1)$}
\\ \toprule
\end{tabular}\label{tab:SMProjection}
\end{center}
\end{table}

The representation content is completed by the vector bosons, whose physical eigenstates include five exotic ones implied by the extended symmetry. The gauge eigenstates enter the kinetic Lagrangian of the theory as

\begin{equation}
\begin{split}
\mathcal{L}^{331}_\mathrm{kin} &= \sum_\Psi i \bar{\Psi} \gamma^\mu D^\mu \Psi + \sum_s (D_\mu s)(D^\mu s) - \frac{1}{4} g_{\mu \nu}^a g^{a \mu \nu} - \frac{1}{4} W_{\mu \nu}^a W^{a \mu \nu} - \frac{1}{4} B_{\mu \nu} B^{ \mu \nu},
\end{split}
\end{equation}
where $D_\mu = \partial_\mu - i g_{s}\frac{\lambda^a}{2} g_\mu^a - i g_{3L}T^a W_\mu^a-i g_X X B_\mu$, and $T^a = \frac{\lambda^a}{2}$ for triplets and $-\frac{\lambda^{aT}}{2}$ for anti-triplets. As for notation, $\Psi$ labels every fermion and $s$ every scalar multiplet, $g^a$ denote the gluons, $W^a$ the $SU(3)_L$ and $B_\mu$ the $U(1)$ gauge bosons. 

After diagonalization of this spin-1 sector, the electrically charged sector is comprised by the $W$ and new singly and doubly charged bosons, written in terms of the symmetry eigenstates simply as

\begin{equation}
\begin{split}
W_\mu^\pm&=(W_\mu^1\mp iW_\mu^2)/\sqrt{2} \\
V_\mu^\pm&=(W_\mu^4\pm iW_\mu^5)/\sqrt{2} \\
U_\mu^{\pm\pm}&=(W_\mu^6\pm iW_\mu^7)/\sqrt{2}, \\
\end{split}
\end{equation}
with masses

\begin{equation} \label{eq:chargedbmass}
\begin{split}
M_W^2&=\frac{1}{4}g_{3L}^2 (v_\eta^2 + v_\rho^2 + 2v_s^2) \equiv \frac{1}{4}g_{3L}^2 v_W^2 \\
M_V^2&=\frac{1}{4}g_{3L}^2( v_\eta^2 + 2v_s^2+v_\chi^2)\\
M_U^2&=\frac{1}{4}g_{3L}^2( v_\rho^2 + 2v_s^2+v_\chi^2).
\end{split}
\end{equation}
Note that, to numerically fit the $W$ mass, $v_\eta^2 + v_\rho^2 + 2v_s^2$ must sum to the Higgs VEV $v_W^2$.

The last addition to the gauge sector with respect to the SM is an extra neutral vector boson $Z^\prime$. The diagonalization of the neutral sector is immensely more complicated and is given, in general, by

\begin{equation}\label{eq:bsymmass}
\begin{split}
W^3_\mu&=-N_1 a_1 Z_{1\mu}-N_2 a_2 Z_{2\mu} + \frac{t_X}{\sqrt{4t_X^2+1}}A_\mu \\
W^8_\mu&= -\sqrt{3}N_1 b_1 Z_{1\mu} -\sqrt{3}N_2 b_2 Z_{2\mu} -\frac{\sqrt{3}t_X}{\sqrt{4t_X^2+1}}A_\mu \\
B_\mu&=2t_X\left(1-\bar{v}_\rho^2\right)N_1 Z_{1\mu} +  2t_X(1-\bar{v}_\rho^2)N_2 Z_{2\mu} + \frac{1}{\sqrt{4t_X^2+1}}A_\mu, \\
\end{split}
\end{equation}
with

\begin{equation}
\begin{split}
a_{1(2)} &=  3 m_{2(1)}^2+\bar{v}_\rho^2-2\bar{v}_W^2 \\
b_{1(2)} &= m_{2(1)}^2+\bar{v}_\rho^2-\frac{2}{3}\bar{v}_W^2-\frac{2}{3}, 
\end{split}
\end{equation}
where the overbar indicates the ratio by $v_\chi$ as $\bar{v}_\alpha\equiv \frac{v_\alpha}{v_\chi}$, and $t \equiv \tan \theta_X \equiv \frac{g_X}{g_{3L}}$. The normalization factors are given by

\begin{equation}\label{eq:normfactors}
\begin{split}
N_1^{-2}&=3 \left(2m_2^2+\bar{v}_\rho^2-\frac{4}{3}\bar{v}_W^2-\frac{1}{3}\right) + (\bar{v}_\rho^2-1)^2(4t_X^2+1) \\
N_2^{-2}&=3 \left(2m_1^2+\bar{v}_\rho^2-\frac{4}{3}\bar{v}_W^2-\frac{1}{3}\right) + (\bar{v}_\rho^2-1)^2(4t_X^2+1).
\end{split}
\end{equation} 
We have defined the factors

\begin{equation}\label{eq:massfactors}
\begin{split}
A&=\frac{1}{3}\left[ 3t_X^2\left(\bar{v}_\rho^2+1 \right) + \bar{v}_W^2 + 1\right] \\
R&=\left\lbrace 1-\frac{1}{3A^2}\left(4t_X^2+1 \right)\left[\bar{v}_W^2\left(\bar{v}_\rho^2+1 \right) -\bar{v}_\rho^4\right]\right\rbrace^{1/2},
\end{split}
\end{equation}
and the dimensionless masses 

\begin{equation} \label{eq:neutralbmass}
\begin{split}
m_{1}^2&=\frac{2 M^2_{Z_1}}{g_{3L}^2 v_\chi^2}=A(1-R) \\
m_{2}^2&=\frac{2 M^2_{Z_2}}{g_{3L}^2 v_\chi^2}=A(1+R).
\end{split}
\end{equation}

This rotation is highly dependent on the vacuum expectation values (VEVs) $v_\rho$ and $v_\chi$. Now, $v_\rho$, $v_\chi$, $v_\eta$ and $v_s$ are free parameters of the theory and their skeptical analysis is out of the scope of this work. Here, when needed, we adopt a benchmark resulting from demanding that the 3-3-1 neutral current parameters reproduce those of the SM ones at the electroweak scale.

This \textit{solution to the closure} we adopt was found by Dias et al~\cite{Dias:2006ns}, who showed that imposing 

\begin{equation} \label{eq:soldef}
v_\rho^2=\frac{1-4s_W^2}{2c_W^2}v_W^2,
\end{equation}
is enough to guarantee that both neutral current parameters, known vector boson masses and the custodial symmetry $\rho$-parameter collapse to their expected values.  Numerically, the solution implies $v_\rho \approx \SI{54}{GeV}$, which, together with the fitting of the $W$-mass in Eq.~(\ref{eq:chargedbmass}), implies $v_\rho \approx \SI{240}{GeV}$, where $v_s$ has been assumed negligible.

The mass matrices of the exotic quarks are of the form

\begin{equation}
M^J_{ab} = \frac{v_\chi}{\sqrt{2}}Y^J, \;\;\;\; M^{j_i}_{ab} = \frac{v_\chi}{\sqrt{2}}Y^{j_i}_{ab},
\end{equation}
which are proportional to the large $v_\chi$ and can be made arbitrarily massive by the free (besides possible phenomenological and unitarity constraints) Yukawa couplings $Y^J,Y^{j_i}_{ab}$. As for the scalars, the model contains, in total, four singly-charged, three doubly-charged, five $\mathit{CP}$-even and three $\mathit{CP}$-odd neutral physical scalars. For the minimization of the most general potential and subsequent diagonalization of the various mass matrices, see~\cite{PhysRevD.69.095009,Tully:1998wa}. The sheer amount of exotic parameters in the scalar sector together with the absence of TeV scale phenomenological constraints allow exotic scalar masses to be large as well. In this work, the words {\it exotic} and {\it heavy} will be used interchangeably to qualify all the particles in the set

\begin{equation}
\left\{ Z^\prime, \; V^\pm, \; U^{\pm \pm}, \; j_1^{-4/3}, \; j_2^{-4/3}, \; J^{5/3}, \; \text{15 exotic scalars}\right\}.
\end{equation}

\section{Approximating the 3-3-1 by the SM symmetry}
\label{sec:ext}

\subsection{Matching and strategy}
\label{sec:extMath}

The usual manner to investigate the running structure of the gauge couplings takes advantage of the fact that, below the scale of importance of the exotic particles (all heavier), the 3-3-1 Model must be approximated by the SM. This is a consequence of the natural decoupling of the heavier particles and of the apparent absence of new physics up to the TeV scale. Furthermore, the SSB of the model automatically splits into two processes, a fact which can be used to describe the approximation mentioned above at a given energy range. From this perspective, the breaking pattern may be written as

\begin{equation} \label{eq:extbreaking}
\begin{split}
\hspace*{-10cm} SU(&3)_c \times SU(3)_L \times U(1)_X \\
&\hspace*{8mm} \xrightarrow[E_\text{high}]{\langle \chi \rangle} SU(3)_c \times SU(2)_L \times U(1)_Y  \\
&\hspace*{11mm}\hspace{10mm}\xrightarrow[E_\text{low}]{\langle \eta \rangle,\langle \rho \rangle} SU(3)_c \times U(1)_{\text{EM}}.
\end{split}
\end{equation}
The schematics above represents that, at the unknown $E_\text{high}$ scale, the 3-3-1 symmetry is broken down to the SM one, after which, at $E_\text{low}$ (to be identified with the electroweak scale), the breakdown to the conserved sector occurs.

Referring to this, the full process induced by the RGE transformations is understood as follows: (\textit{i}) Nature is assumed to be well described by the $SU(3)_c \times SU(2)_L \times U(1)_Y$ symmetry immediately above an $E_\text{low}$ scale, below which the theory is supposed to be broken to the conserved $SU(3)_c \times U(1)_\text{EM}$; At this threshold, the couplings are fixed to their numerically known values; (\textit{ii}) $g_{2L},g_{Y}$ are evolved, according to the appropriate particle content (see below and in the next section), with increasing energy, up to $E_\text{high}$; (\textit{iii}) The second matching is performed to replace $g_{2L},g_{Y}$ by the emergent $g_{3L},g_{X}$; (\textit{iv}) The couplings are evolved again, now up to the pole, according to a second structure corresponding to the higher symmetry and larger set of particles suitable to the new energy range. Notice that $E_\text{high}$ is the scale in which the effects of the heavy particles become important, which causes the apparent symmetry to change.

At $E_\text{low}$, generically taken to be at the $Z$-pole from now on, $E_\text{low} = M_Z \approx \SI{91.2}{GeV}$, the known SM couplings are simply fixed to their well measured numerical values \cite{ParticleDataGroup:2022pth}:

\begin{equation}
g_{2L}(E_\text{low}) = 0.630, \;\;\;\;\; g_Y(E_\text{low}) =  0.345.
\end{equation}

\begin{table}[t!]
\caption{$\beta$-function coefficients of the m331 couplings in the two energy ranges. Two configurations for the representation content are emphasized: `Full' refers to the entire 3-3-1 content, whereas only the SM degrees of freedom contribute to `SM'.}
\begin{center}
\begin{tabular}{ccccc}
\toprule
\hphantom{$E_\text{low}<\mu<E_\text{high}$} & $g_{Y(X)}$: Full & $g_{2L(3L)}$: Full & $g_Y$: SM & $g_{2L}$: SM
 \\ \midrule
$E_\text{low}<\mu<E_\text{high}$  &   $38$      &  $-2$    &  41/6   &  -19/6 \\
$\mu>E_\text{high}$              &   $22$      &  $-17/3$ &  &   \\ \toprule
\end{tabular}\label{Tab:bcoeffext}
\end{center}
\end{table}

At $E_\text{high}$, the two exotic couplings may be matched to the SM ones according to the formalism reviewed in Sec.~\ref{sec:matching}. Regarding $g_{3L}$, it is enough to realize that the relevant embedding structure obeys $SU(2)_L \subset SU(3)_L$, causing the necessary condition to be simply

\begin{equation}\label{eq:g3Lg2L}
g_{3L}(E_\text{high}) = g_{2L}(E_\text{high}).
\end{equation}

For $g_X$, we recall 

\begin{equation}
Y = -\sqrt{3} T_8 + X,
\end{equation}
which is a consequence of Eq.~(\ref{eq:charge}). Of inserting this information into formula (\ref{eq:matchEq}), results

\begin{equation}\label{eq:Matchingextbig}
\frac{1}{g_Y(E_\text{high})^2} = \frac{3}{g_{3L}(E_\text{high})^2} + \frac{1}{g_{X}(E_\text{high})^2},
\end{equation}
or, plugging Eq.~(\ref{eq:g3Lg2L}),

\begin{equation} \label{eq:gXFinalMatching}
g_X(E_\text{high}) = \frac{g_{2L}(E_\text{high}) \, g_Y(E_\text{high})}{\sqrt{g_{2L}(E_\text{high})^2 - 3 g_Y(E_\text{high})^2}}.
\end{equation}

Once more, it must be emphasized that these boundary conditions on the breaking scales are merely consistency requirements to guarantee the supposed symmetry structure and continuity. In fact, they obviously could be found -- and usually are -- by explicit brute force comparison of the two lagrangians with correct symmetry properties around the threshold. 

The $\beta$-function coefficients are straightforward to calculate through Eq.~(\ref{eq:betacoeff}) and are shown in Table~\ref{Tab:bcoeffext}. The values corresponding to the m331 with the exotic degrees of freedom removed are also presented. Note that removing the entire exotic sector within the 3-2-1 (SM symmetry) energy range amounts to eliminating closed multiplets, keeping the theory symmetric. This assures that the exact RGE may be solved directly, with the $\beta$-function coefficients given by the classic result of Eq.~(\ref{eq:betacoeff}).

\subsection{Results}

If the entire particle content is naively kept from very low energies, the explicit $g_X$ dependence on energy is given by

\begin{equation}
\begin{split}
g_X^{\text{full}}(\mu) = 2\pi (3&0.1 - 11.0 \log{E_\text{high}} + 22.0 \log{E_\text{low}} - 11.0 \log{\mu})^{-1/2}, \;\;\;\; \mu>E_\mathrm{high},
\end{split}
\end{equation}
with the immediate result that for $E_\text{high} \gtrsim \SI{370}{GeV}$
$g_X^{\text{full}}$ becomes imaginary in the entire higher range. This signifies that, for these choices of $E_\mathrm{high}$, the mandatory matching conditions are impossible to fulfil, rendering the theory senseless in the higher symmetry regime. The upper bound on $E_\text{high}$ then reads

\begin{equation}
E_\text{high} \lesssim \SI{370}{GeV}.
\end{equation}
However, such small values for the breaking scale are clearly not phenomenologically viable. 

\begin{figure*}[t!]
	\centering
		\includegraphics[width=0.7\linewidth]{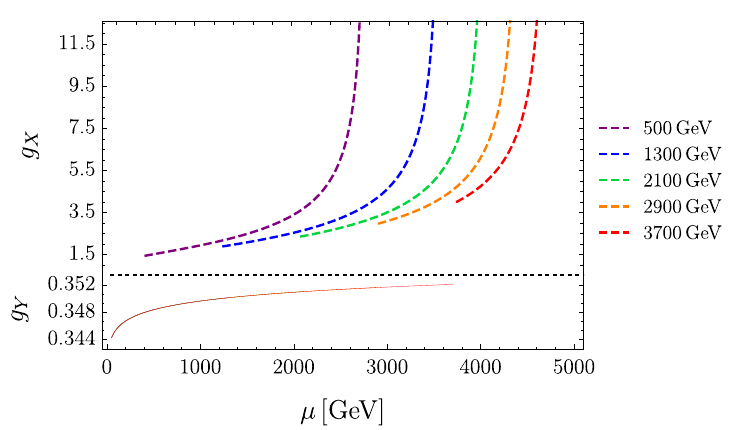}
		\caption{\textit{Extension scenario} results for the running of $g_Y$ (solid) and $g_X$ (dashed) for five different $E_\text{high}$ benchmarks.}
	\label{fig:extSMdof}
\end{figure*}

This result does not come as a surprise, since in $\overline{\text{MS}}$ the decoupling of heavy particles is not automatic as happens in physical renormalization schemes, and must be introduced by hand \cite{Appelquist:1974tg}. If enforced, and only the known particles are kept at energies $\mu < E_\text{high}$, with the heavy degrees of freedom integrated out and only included above this threshold, the coupling depends on energy as

\begin{equation}
\begin{split}
g_X^{\text{SM}}(\mu) = 2\pi (3&0.1 + 2.83 \log{E_\text{high}} + 8.17 \log{E_\text{low}} - 11.0 \log{\mu})^{-1/2}, \;\;\;\; \mu>E_\mathrm{high}.
\end{split}\label{eq:gXknown}
\end{equation}
The corresponding behaviour for several benchmark $E_\text{high}$ is shown in Figure~\ref{fig:extSMdof}. It may be observed that, for breaking scales obeying $E_\text{high} \gtrsim \SI{3770}{GeV}$, the $g_X$ coupling is highly vertical, originating larger than $4 \pi$ and rapidly diverging, and there is no effectively perturbative window. The upper bound with the heavy particles decoupled becomes

\begin{equation}
E_\text{high} \lesssim \SI{3770}{GeV}.
\end{equation}
Unfortunately, even for the optimal $E_\text{high} = \SI{3770}{GeV}$ perturbativity is quickly harmed at around $\sim \hspace*{-0.4mm}\SI{4600}{GeV}$. This result is usually quoted without the corresponding choice of $E_\text{high}$, which, in turn, is usually thought to define the $v_\chi$ parameter. In summary, assuming the position for the pole in $g_X$ automatically induces phenomenological consequences.

The conclusions of this section could be qualitatively understood by noting that, in the matching condition,

\begin{equation}
\frac{1}{g_X^2} =  \frac{1}{g_Y^2}-\frac{3}{g_{2L}^2},
\end{equation}
the RHS is smaller than one ($\sim \hspace*{-0.4mm}0.84$) already at the electroweak scale, and its reciprocal very sensitive to small changes in $g_Y$ and $g_{2L}$, which are increasing and decreasing, respectively.  

\section{Effective approach}
\label{sec:alt}

\subsection{Matching and strategy}
\label{sec:altMath}

The conclusion from the last section is that, to match an exact 3-2-1 -- the SM as embedded into the m331 -- to an exact 3-3-1 theory is only possible in a very low regime. Such attempts are justified by the reasoning that, if the exotic particles are ignored, interactions should be approximately parametrized as within the SM. This, in turn, should be possible within a model like the m331 because they are constructed with an SM embedding in mind, which is a clear way to guarantee that its good predictions are reproduced. To use the SM as a low energy description of an extension, however, is only an approximation (to be made around a given characteristic energy scale), and does not translate well to an RGE analysis. At zero temperature, all that matters are the form and content of the non-decoupled interactions, and the definition of an active intermediate symmetry, as in Eq.~(\ref{eq:extbreaking}), is only a convenient shortcut to describe them in a theory with the embedded SM and every exotic effect decoupled. This, however, is not necessarily possible at an arbitrary mass regime. Note, in this sense, that in a general model it is not strictly necessary for the SSB pattern to decompose as in the m331: the VEV acquisition for distinct scalars (or subgroups of them) could each trigger the direct breaking to the electromagnetism or to arbitrary intermediate groups. Nonetheless, in principle, such a theory could still produce all the electroweak predictions of the SM.    

We shall now take on a more precise approach, in which the 3-3-1 parametrization is assumed from the start. What can be gained from this is the circumventing of the matching process (which is artificial and an approximation), and the downside is the need for a brute force, explicit approach, as Eq.~(\ref{eq:betacoeff}) is no longer valid when the heavy degrees of freedom are integrated out. 

To describe the coupling evolution process, notice that $g_{2L}$ and $g_Y$ are now undefined, and $g_{3L}$ and $g_X$ hold unrelated new values, to be numerically matched at $E_\text{low}$ with experimentally obtained quantities. For that,
match the strength of the $\gamma e \bar{e}$ and $W e \bar{\nu}$ interactions to their experimental counterpart. The requirements read

\begin{equation}\label{eq:EffMatching}
\begin{split}
\Gamma_\mu^{\gamma e \bar{e}} &= i\gamma_\mu \left(   -\frac{g_{3L}}{2} \mathcal{O}_{13}P_L + \frac{g_{3L}}{2\sqrt{3}} \mathcal{O}_{23}P_L + \frac{g_{3L}}{\sqrt{3}} \mathcal{O}_{23}P_R   \right) \overset{!}{=} -i|e|\gamma_\mu \\
\Gamma_\mu^{W e \bar{\nu}} &= i \frac{g_{3L}}{\sqrt{2}} \gamma_\mu \overset{!}{=} i \frac{g_{2L}}{\sqrt{2}} \gamma_\mu
\end{split}
\end{equation}
where $\mathcal{O}$ rotates mass to symmetry eigenstates like $(W_{3\mu},W_{8\mu},B_{\mu})^T = \mathcal{O}\cdot (Z_{\mu},Z^\prime_{\mu},A_\mu)^T$, and is given explicitly in Eq.~(\ref{eq:bsymmass}). In the end, the matching amounts to setting

\begin{equation}
g_{3L}(M_Z) = 0.63, \;\;\;\; g_{X}(M_Z) = 1.1.
\end{equation}

In order to understand how heavy particle decoupling may be implemented, recall that, in the last section, the theory that remained after elimination of the heavy particles continued symmetric in the SM approximation, and it was enough to modify the $\beta$-function coefficients to encompass the light multiplets alone. This was made possible because the SM projection of the m331 contains every exotic degree of freedom as a singlet, which may be removed without spoiling the symmetry. An analogous situation occurs if one inspects the RGE of the SM without the third generation of quarks. In the other hand, if the gauge symmetry is broken by disregarding a few particles, some brute force method becomes necessary. This is exemplified by the exclusion of the top quark from the SM, and corresponds to our problem at hand since the $SU(3)_L$ triplets and octet are broken once the heavy particles are removed. The theory is no longer symmetric below $E_\mathrm{high}$, and an explicit, specific calculation cannot be escaped. 

The procedure amounts to subtracting from the complete, symmetric $\beta$-functions, the part resulting from the 1-loop contributions of the exotic particles (indicated by an $\mathscr{H}$ superscript), $\delta Z^\mathscr{H}_{g_{3L}}$ and $\delta Z^\mathscr{H}_{g_X}$, only activated above the $E_{\mathrm{high}}$ threshold. For that end, we construct a completely determined system of equations from which the counterterms of the couplings may be obtained. In order to build such a system, the counterterms for at least two vertex functions are required, and are conveniently chosen to be the $W u \bar{d}$ and $\gamma u \bar{u}$ ones. 

To summarize, the evolution process within the current construction follows as:  (\textit{i}) At $E_\text{low}$, we match $g_{3L}$ and $g_X$ to fit $Z$-pole interactions as in Eq.~(\ref{eq:EffMatching});  (\textit{ii}) We initially evolve the couplings with increasing energy according to an effective representation content where every exotic particle has been integrated out through some procedure; (\textit{iii}) At a parametrically free scale $E_\text{high}$, referring now to the one in which their effects become important, we include the heavy particles back into the theory, match to the model below, and evolve the couplings up to the pole. Notice that only $SU(3)_L \times U(1)_X$ quantities are ever mentioned. 

\subsection{Renormalization framework}

We renormalize the various quantities by making\footnote{Note the distinction between our convention and the common one which renormalizes the gauge vertex as $\Gamma_0^{V f_1 \bar{f_2}} = Z_g \Gamma^{V f_1 \bar{f_2}}$. In that alternative definition, one has $g_0 = \frac{Z_g}{Z_V^{1/2} Z^{1/2}_{f_1}Z^{1/2}_{f_2}}g$~\cite{Srednicki:2007qs}.}

\begin{equation}
f_0 = Z_f^{1/2} f, \;\;\;\; V_0 = Z_V^{1/2} V, \;\;\;\; g_0 = Z_g g,
\end{equation}
where $f$ is any fermion, $V$ any vector boson field and $g$ any gauge coupling. The vertex functions to be renormalized read

\begin{equation}\label{eq:verticesRen}
\begin{split}
\Gamma_{m_1 m_2 \mu}^{W u \bar{d}} & = i \frac{g_{3L}}{\sqrt{2}} \delta_{m_1 m_2} \gamma_\mu P_L \\ 
\Gamma_{m_1 m_2 \mu}^{\gamma u \bar{u}} &= i\frac{g_{3L}}{2}\delta_{m_1 m_2} \gamma_\mu \left[ \left( \mathcal{O}_{13} - \frac{\mathcal{O}_{23}}{\sqrt{3}} - \frac{2\mathcal{O}_{33}}{\sqrt{3}} \right)P_L + \frac{4 t_X \mathcal{O}_{33}}{3}P_R \right],
\end{split}
\end{equation}
where $m_1,m_2$ are color indices. 

At this point, in order to solve for $\delta Z^\mathscr{H}_{g_{3L}}$ and $\delta Z^\mathscr{H}_{g_{X}}$, the set of functions 

\begin{equation}
\left\{\delta \Gamma^\mathscr{H}_{W u\overline{d}},\delta \Gamma^\mathscr{H}_{\gamma u\bar{u}},\delta Z^\mathscr{H}_{u},\delta Z^\mathscr{H}_{d},\delta Z^\mathscr{H}_{W},\delta Z^\mathscr{H}_{A}\right\},
\end{equation}
must be evaluated. Since the dependence of all these quantities on their individual 1-loop contributions is additive, they may be calculated referring only to the exotic diagrams, shown in Fig.~\ref{fig:diagrams}, and there are no crossed effects between those and the contributions of the pure SM. From Eq.~(\ref{eq:verticesRen}), the relations among the vertex function counterterms $\delta \Gamma^\mathscr{H}$ and the other ones may be found and read (omitting the Lorentz and color indices in the LHS for clarity)

\begin{equation}\label{eq:counterterms}
\begin{split}
\delta \Gamma^\mathscr{H}_{W u \bar{d}} &= i  \frac{g_{3L}}{2\sqrt{2}}\delta_{m_1 m_2} \gamma_\mu P_L\left( 2\delta Z^{\mathscr{H}}_{g_{3L}} + \delta Z^{\mathscr{H}}_{d} + \delta Z^{\mathscr{H}}_{u} + \delta Z^{\mathscr{H}}_{W} \right)\\
\delta \Gamma^\mathscr{H}_{\gamma u \bar{u}} &= i \frac{\gamma_\mu}{2}\delta_{m_1 m_2}  \left[g_{3L}P_L \left( \mathcal{O}_{13}-\frac{\mathcal{O}_{23}}{\sqrt{3}}\right)\delta_\mathrm{E}Z^{\mathscr{H}}_{g_{3L}}  \right. \\ 
& \left. {} \hspace*{22mm} + g_{X} \left( \frac{4\mathcal{O}_{33}}{3}P_R -\frac{2\mathcal{O}_{33}}{\sqrt{3}}P_L \right)\delta_\mathrm{E}Z^{\mathscr{H}}_{g_{3L}}  \right],
\end{split}
\end{equation}
where we have defined the `effective' coupling counterterms

\begin{equation}
\begin{split}
\delta_\text{E} Z^{\mathscr{H}}_{g_{3L}}&\equiv \delta Z^\mathscr{H}_{g_{3L}} + \delta Z^{\mathscr{H}}_u + \frac{\delta Z^{\mathscr{H}}_A}{2} \\
\delta_\text{E} Z^{\mathscr{H}}_{g_{X}}&\equiv\delta Z^\mathscr{H}_{g_{X}} + \delta Z^{\mathscr{H}}_u + \frac{\delta Z^{\mathscr{H}}_A}{2},
\end{split}
\end{equation}
where $Z_{x} \equiv 1 + \delta Z_x$.

The solution of this system is then used to find the $\beta^\mathscr{H}$-function by demanding perturbative consistency from the RGE $\partial g_0 / \partial \ln \mu = 0$ as usual \cite{Weinberg:1996kr}. The running is finally obtained, in the non-symmetric regime below the heavy particles threshold $E_\mathrm{high}$, by numerically solving the coupled system of differential equations 

\begin{align}
\frac{\partial g_{3L}(\mu)}{\partial \ln \mu} - \frac{g_{3L}(\mu)^3}{(4 \pi)^2}b_{g_{3L}} &- \left[- \beta^\mathscr{H}_{g_{3L}}\left(g_{3L}(\mu),g_{X}(\mu)\right) \right] = 0, \label{eq:NSolving1} \\
\frac{\partial g_{X}(\mu)}{\partial \ln \mu} - \frac{g_{X}(\mu)^3}{(4 \pi)^2}b_{g_{X}} &- \left[- \beta^\mathscr{H}_{g_{X}}\left(g_{3L}(\mu),g_{X}(\mu)\right) \right] = 0. \label{eq:NSolving2}
\end{align}

An important remark is that the exotic scalars are not included in the calculation of the counterterms (see Fig.~\ref{fig:diagrams}). This is because they come in fully exotic triplets, and can be correctly eliminated from the theory through the subtraction of their contributions to $b_{g_X},b_{g_{3L}}$. The exception is the SM scalar doublet, which must be kept. We consider it to be the one projected by the $\rho$-triplet, $\rho^{(2)}$  from Table~\ref{tab:SMProjection}, and conserve its contribution to the coefficients. The $b_{g_X},b_{g_{3L}}$ to be plugged into the symmetric term of the equations above, already accounting for the removal of the scalars other than $\rho^{(2)}$, are

\begin{equation}
b_{g_{X}} = \frac{62}{3}, \;\;\;\;\; b_{g_{3L}} = -\frac{41}{6}.
\end{equation}  

\subsection{Results}
\label{sec:resAlt}

The objective is to investigate whether the more precise effective approach can alleviate the stress that the $\SI{4}{TeV}$ pole, obtained through an approximation that makes use of an artificial intermediate exact symmetry, generates over the model. The operational method implies several steps, and many amplitudes of Figure~\ref{fig:diagrams} are derived from rules which originate from Lagrangians with explicit charge conjugation, which turns their obtaining into a subtle matter \cite{Denner:1992vza,Barela:2022sbb}. We check them using \texttt{FeynRules} \cite{Alloul:2013bka} paired with \texttt{FeynArts} \cite{Hahn:2000kx}. The loop calculations are performed with the help of \texttt{Package-X} \cite{Patel:2016fam} connected to \texttt{FeynCalc} through \texttt{FeynHelpers} \cite{Shtabovenko:2016whf}. The Feynman-'t Hooft gauge \cite{Fujikawa:1972fe} is employed when needed.

In order for Eqs.~(\ref{eq:NSolving1}) and~(\ref{eq:NSolving2}) to be solved for $g_X$ below $E_\mathrm{high}$, the quantities $v_\rho$, $v_\eta$, and $v_\chi$ must be fixed. We use the {\it solution to the closure} mentioned near Eq.~(\ref{eq:soldef}) to set $v_\rho = \SI{54}{GeV}$ and $v_\eta = \SI{240}{GeV}$.Additionally, to fix the values of the exotic quarks masses, we set their three Yukawa eigenvalues to $0.5$, {\it i.e.},

\begin{equation}
Y^J = 0.5, \;\;\;\; Y^{j} = 0.5 \,\mathds{1}.
\end{equation} 
With this, we examine the following four benchmarks for $v_\chi$, with every exotic fermionic and spin-1 mass defined in consequence: 

\begin{equation}\label{eq:Benchmarks}
\begin{split}
&\mathrm{B1}\hspace{-0.1cm}: \quad v_\chi=\SI{3}{TeV}, \, M_U = \SI{945}{GeV}, \, M_V = \SI{948}{GeV}, \, M_{Z^\prime} = \SI{3476}{GeV},\\
&\hphantom{\mathrm{B1}\hspace{-0.1cm}: \quad v_\chi=\SI{3}{TeV},} \hspace*{2mm} M_j = \SI{1061}{GeV}, \,M_J = \SI{1061}{GeV};   \\[0.22em]
&\mathrm{B2}\hspace{-0.1cm}: \quad v_\chi=\SI{6.5}{TeV}, \, M_U = \SI{2048}{GeV}, \, M_V = \SI{2049}{GeV}, \, M_{Z^\prime} = \SI{7531}{GeV},\\
&\hphantom{\mathrm{B1}\hspace{-0.1cm}: \quad v_\chi=\SI{6.5}{TeV},} \hspace*{2mm} M_j = \SI{2298}{GeV}, \,M_J = \SI{2298}{GeV};   \\[0.22em]
&\mathrm{B3}\hspace{-0.1cm}: \quad v_\chi=\SI{9.5}{TeV}, \, M_U = \SI{2993}{GeV}, \, M_V = \SI{2993}{GeV}, \, M_{Z^\prime} = \SI{11}{TeV},\\
&\hphantom{\mathrm{B1}\hspace{-0.1cm}: \quad v_\chi=\SI{9.5}{TeV},} \hspace*{2mm} M_j = \SI{3359}{GeV}, \,M_J = \SI{3359}{GeV};   \\[0.22em]
&\mathrm{B4}\hspace{-0.1cm}: \quad v_\chi=\SI{13}{TeV}, \, M_U = \SI{4095}{GeV}, \, M_V = \SI{4096}{GeV}, \, M_{Z^\prime} = \SI{15.06}{TeV},\\
&\hphantom{\mathrm{B1}\hspace{-0.1cm}: \quad v_\chi=\SI{13}{TeV},} \hspace*{2mm} M_j = \SI{4596}{GeV}, \,M_J = \SI{4596}{GeV}.    
\end{split}
\end{equation}
B1, the most conservative benchmark point, is chosen with $M_U \approx \SI{1}{TeV}$ because this is a very conservative reasonable lower bound for the mass of the bilepton given by the joint phenomenology already produced for this particle~\cite{Meirose:2011cs,Nepomuceno:2016jyr,Corcella:2017dns,RamirezBarreto:2013edk,PhysRevD.101.015024,Barela:2022sbb}.  

Besides $v_\chi$, $E_\mathrm{high}$ is yet to be fixed. This parameter is directly identifiable with the scale of importance of the heavy states. In each of the Benchmarks above, we set it to the mass of the vector bilepton $E_\mathrm{high}=M_U$ which, except for $M_{Z^\prime}$, corresponds to a good representative for the scale of all masses. An even more complete assessment of this process would integrate each heavy state out at their exact mass point (or, e.g., at a predefined fraction of it). Note that, in general, there is still a loose relation between a `characteristic scale' of the higher symmetry, $E_\mathrm{high}$ and $v_\chi$, since the latter is strongly connected to exotic masses. 

To validate our calculations, we derive the runnings through a second, approximate method of integrating the exotic particles out below $E_\mathrm{high}$. The procedure amounts to, as before, modifying the $b$ in an attempt to remove their effects. In principle, this could be done exactly for $g_X$, as each contribution is computed from a single particle. However, because there is mixing between the $U(1)_X$ and the `diagonal' gauge bosons of $SU(3)_L$, the results could, in principle, differ, mainly because of the physical $Z^\prime$. This could also
play a role in the form and strength of the interactions, another possible source of imprecision. The situation is more critical for the $g_{3L}$ running, since the quark triplets are broken by the removal of $j_i,J$. To find $b_{g_{3L}}$, we ignore this fact, and consider the contribution of the $S_2$ of each broken triplet to Eq.~(\ref{eq:betacoeff}) as if they were intact --  we dub this method {\it doublet approximation}. Note that the $\beta$-function of $g_X$ depends on $g_{3L}$, hence, again, although this procedure could be used expecting exact results for $g_X$, there could also be distortions caused by the $g_{3L}$ error. The $\beta$-function coefficients to be used in the $\mu< E_\mathrm{high}$ regime in this approximation are

\begin{figure*}[t!]
	\centering
	\hspace*{-0.11cm}\makebox[0pt][c]{%
		\includegraphics[width=0.5\linewidth]{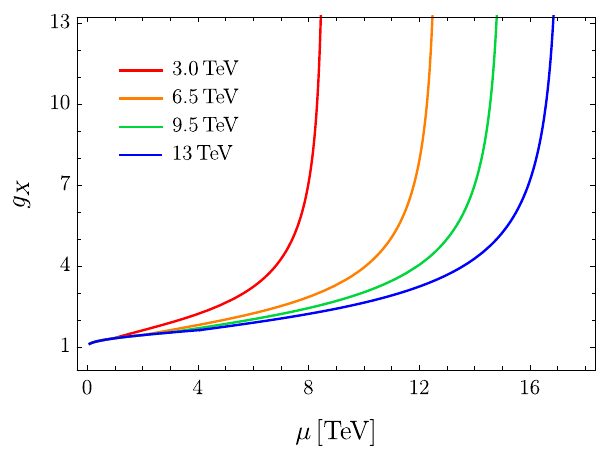}
		\includegraphics[trim=0 0.5mm 0 0, clip, width=0.52\linewidth]{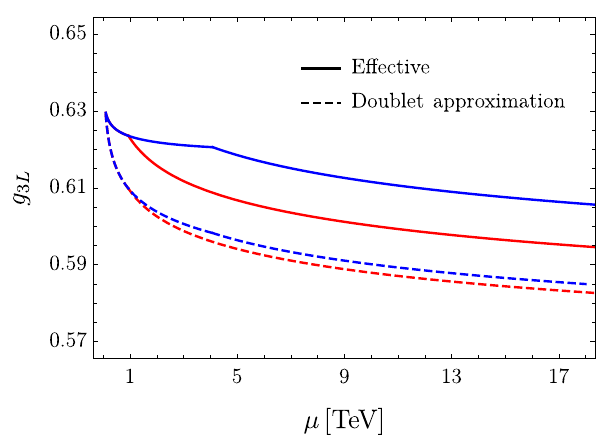}}
		\caption{\textit{Left}: Effective (exact) evolution of the running of $g_X$ for the four benchmarks defined in Eq.~(\ref{eq:Benchmarks}), labeled by the free $v_\chi$. The plot attests that avoiding the SM approximation greatly enlarges the perturbativity regime of the model, depending on $v_\chi$. \textit{Right}:  Evolution of the left-handed 3-3-1 coupling (to avoid cluttering in the image, only the two extreme benchmarks appear). The dashed line corresponds to the running obtained through the {\it doublet approximation}.}
	\label{fig:altSMdof}
\end{figure*}

\begin{equation}
b_{g_{X}} = \frac{55}{9}, \;\;\;\;\; b_{g_{3L}} = -\frac{17}{3}.
\end{equation}

\begin{figure*}[t!]
	\centering
	\hspace*{-0.11cm}\makebox[0pt][c]{%
		\includegraphics[width=0.5\linewidth]{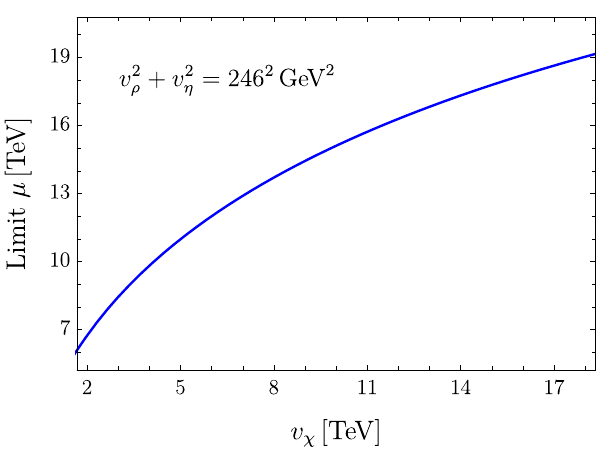}
		\includegraphics[trim=0 0.5mm 0 0, clip, width=0.5\linewidth]{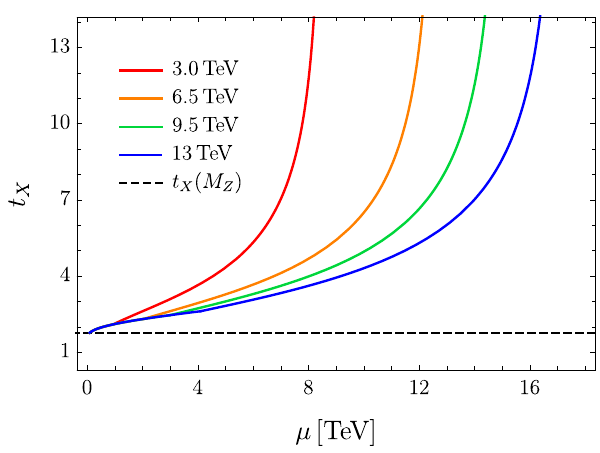}}
		\caption{{\it Left}: Upper limit of the perturbative range of the m331 as a function of $v_\chi$. This curve is independent of $v_\rho,v_\eta$ as long as they belong to the circle $v_\rho^2+v_\eta^2 = \SI{246^2}{GeV^2}$, with $v_s \approx \SI{2}{GeV}$ (negligible in practice). {\it Right}: Running of the m331 symmetry parameter $t_X \equiv g_X/g_{3L}$ for our four benchmark points, to be compared with the usually employed value $t_X = t_X(M_Z) \approx 1.75$.}
	\label{fig:polePosition}
\end{figure*}

The results are shown in Figure~\ref{fig:altSMdof}. The first observation to be made is that, despite the discussion carried in the last paragraph, the {\it doublet approximation} is, for all purposes, perfect for the $g_X$ running, i.e., the curves coincide. One way to explain this is to realize that the the greatest source of `mixing' between $g_X$ and $g_{3L}$ comes from $Z^\prime$ effects, whose scale is exaggeratedly larger than $E_{\mathrm{high}}$. Hence, in the scales of matching between the theory of light states and the complete model, such effects are negligible. The greatest result, however, is that considering the rightful parametrization of the model from very low energies extends its unitary range from around $\SI{4.5}{TeV}$ (in the SM approximation) to $\sim \hspace*{-0.4mm} \SI{8.5}{TeV}$ in the most conservative benchmark. This range can be enlarged further with an increasing $v_\chi$ which, besides influencing the neutral spin-1 particles projection onto low energies, has as major consequence pushing the heavy particle threshold upwards. The left panel of Figure~\ref{fig:polePosition} shows the upper limit of the perturbative window of the model, depicted as the energy scale in which $g_X(\mu) = 4 \pi$, as a function of $v_\chi$. An interesting fact that we have verified is that this figure is not altered by a change in $v_\rho$ and $v_\eta$, for fixed $v_\chi$, at least as long as they obey $v_\rho^2 + v_\eta^2 = \SI{246^2}{GeV^2}$.    

Finally, let us try to estimate a measure of the harm of generating TeV scale predictions, in the m331, without conducting RGE improved calculations. In general, the amplitudes for physical processes may be written in terms of $g_{3L}$ and the symmetry parameter $t_X = g_X/g_{3L}$. The right panel of Figure~\ref{fig:polePosition} shows the running of $t_X$ for our four benchmarks. In most 3-3-1 studies, this parameter is eliminated in favour of the known quantity $s_W = g_{2L}/\sqrt{g_{2L}^2+g_{Y}^2}$ through the relation

\begin{equation} \label{eq:tanX}
t_X^2 \equiv \frac{g_X}{g_{3L}} = \frac{s_W^2}{1-4 s_W^2},
\end{equation}
which is nothing but a matching condition between a 3-3-1 and the 3-2-1 SM. As such, it is required to hold at a single energy point (which, in fact, is $E_\mathrm{high}$, and not in the electroweak scale), not as an identity between functions of $\mu$. As this paper has shown, even if one runs $s_W$ with energy, the SM approximation and the artificial matching turn out to be a great source of inaccuracy. In any case, to get a sense of the effects of this disparity, consider the exotic $Z^\prime$ mediated contribution to the hard partonic process $u\bar{u} \to Z^\prime \to e^+e^-$. The lowest order term in $m_u/\hat{s}$ of the averaged, angular inclusive, cross section is given by

\begin{equation}
\hat{\sigma}_{Z^\prime}\left(u\bar{u} \to e^+e^-\right) = \frac{5g_{3L}^4 \hat{s}}{2^8 3^4 \pi \left[\hat{s}^2 + M_{Z^\prime}^4 + M_{Z^\prime}^2(\Gamma_{Z^\prime}^2 - 2\hat{s})\right]} f(\theta_X),
\end{equation}
where $M_{Z^\prime},\Gamma_{Z^\prime}$ are the $Z^\prime$ mass and width, respectively, and $\hat{s}$ is the $u\bar{u}$-pair center-of-mass energy squared. The $f$ factor is given by

\begin{equation}
f(\theta_X) = \frac{60  t_X^6 + 20  t_X^4 + 4\sqrt{3t_X^2 + 1}  t_X^2 + 1}{(3t_X^2 + 1)^2}.
\end{equation}
Now, consider $g_{3L}$ slow varying and a partonic $\hat{s} = \SI{1.5}{TeV}$. In this scenario, one obtains that $\hat{\sigma}_{Z^\prime}$ picks up an extra, wrong factor of $\sim \hspace*{-0.4mm} 1/2$ if one uses $t_X = t_X(M_Z)$, i.e., the value obtained by matching to the SM at the $Z$-pole, the usual practice. This discrepancy increases fast with energy.

\section{Conclusions}
\label{sec:con}

The SSB of the m331 has two groups of contributions: the first, generated by the condensation of the $\chi$-triplet neutral component, triggers the descent of the 3-3-1 symmetry group to that of the SM. The second, originated from every other VEV, prompts the usual SM breaking. This is a mathematical construct of the model building, and its interpretation as a meaningful physical process is a conceptual simplification. Although it is a phenomenological necessity of any BSM theory to possess the SM as an effective approximation below the TeV scale, this does not imply that a 3-2-1 symmetry approximation is appropriate outside a narrow window within the electroweak regime. In fact, to force the SM symmetry as a physical feature of the m331 in intermediate scales is a strong simplification, which becomes strictly impossible for heavy particle thresholds above $\sim \hspace*{-0.4mm} \SI{3.5}{TeV}$, and badly imprecise way before it. We firstly review the prediction of the $\SI{4}{TeV}$ pole in this SM approximation, finding what is the greatest matching scale that leaves a small perturbative range available above it, which corresponds to around $3.7-\SI{3.8}{TeV}$, resulting in a pole at a little above $\SI{4.5}{TeV}$. 

A full account of the most precise, effective approach is then given. We define the heavy particle threshold along the mass of the vector bilepton, one of the most interesting features of the model and which gives a reasonable avatar for the general exotic mass scale. Other important free parameters are the triplet VEVs, $v_\chi$, $v_\rho$, and $v_\eta$, which influence the projection of the neutral vector boson masses and interactions at low energies. The quantities $v_\rho$ and $v_\eta$ are fixed through a numerical solution that fits the known neutral current parameters at the $Z$-pole, and four benchmarks are chosen for $v_\chi$, the lowest of which approximately exhausts the predictions for the lower bounds given by the current bilepton phenomenology. Thus, the full structure of the diagonalization of the neutral sector in the 3-3-1 is taken into consideration, and the exact physical states are removed below their scale of importance.  

Our main results show that, in the most conservative benchmark, coherent with a bilepton mass of $M_U = \SI{945}{GeV}$, the true perturbative range of the 3-3-1 extends up to $\SI{8.5}{TeV}$, already greatly reducing the stress generated onto the model by the usual assessments. For a heavy particle threshold around $M_U = \SI{2990}{GeV}$, a still viable phenomenological (in some sense, more natural~\cite{Barela:2022sbb}) scenario, this window is increased further up to $\SI{15}{TeV}$. Our calculations are validated by reproducing the $\beta$-functions of general, non-abelian theories and, more importantly, by identically matching the {\it doublet approximation} for $g_X$, whose results for this coupling should be highly reliable. This is because the only sources of error in this approach for this interaction come from its mixing with the $SU(3)_L$ one.

Apart from assessing alternative versions of the model, left for posterior works, there are a few points in the analysis, ignored by simplicity, which could be addressed, starting by the possibility of splitting the threshold: we have considered a single heavy particle one. A more thorough analysis of the parametric structure of the model could be carried, considering more realistic, strategically chosen benchmark points which define, for instance, different quark Yukawa couplings and the parameters of the scalar potential. With such a parametric map at hand, one would be able to perform fully realistic RGE analysis, integrating each particle out at their exact mass scale predicted in the given point of parameter space. In particular, a deep assessment of the scalar sector would define the projection of the SM physical Higgs onto the low energies, for which we used the simplest possible benchmark. 

Such new analysis are, however, intrinsically tied to experimental and phenomenological advances and should either not immensely vary the paradigm unveiled here or be tied to distinct and complementary premises. It must be understood that the Landau Poles, by themselves, do not condemn the model, but attest that, at the corresponding energy ranges, new degrees of freedom or theoretical mechanism must arise to protect the theory.

Also of importance is the review of a common practice which eliminates a free electroweak parameter of the model and should ideally be avoided from skeptical investigations. The electroweak angle $t_X \equiv g_X/g_{3L}$ \textit{should not} be equated to $s_W^2/(1-4 s_W^2)$, except if in a conscientious approximated ansatz around a fixed scale. Finally, the importance of our results is not limited to the perturbative qualities of the model, but calls attention to possible effects of modifying the theoretical status of the theory regarding the SSB, which urges RGE improved phenomenology comparing both scenarios to be performed.

\section*{ACKNOWLEDGEMENTS}

The author is grateful to Vicente Pleitez for the idea that originated this project, for many useful discussions throughout it and for the review of this work, and to Rodolfo Capdevilla for helpful conversations and the critical reading of the paper. MB is also grateful to CNPq for the financial support.

\appendix

\section{Contributions of the heavy particles to the $\beta$-function}\label{app:diagrams}

The calculated wavefunction counterterms read

\begin{equation}
\begin{split}
\delta Z^\mathscr{H}_u &= -\frac{g_{3L}^2}{64 \pi^2 \epsilon} \\
\delta Z^\mathscr{H}_d &= -\frac{g_{3L}^2}{64 \pi^2 \epsilon} \\
\delta Z^\mathscr{H}_W &= \frac{5 g_{3L}^2 (2 \mathcal{O}_{12}^2 + 1)}{48 \pi^2 \epsilon} \\
\delta Z^\mathscr{H}_A &= \frac{3 g_{3L}^2 (5 \mathcal{O}_{13}^2 + 9 \mathcal{O}_{23}^2)+ 16\sqrt{3} g_{3L}g_{X} \mathcal{O}_{23}\mathcal{O}_{33}- 126 g_{X}^2 \mathcal{O}_{33}^2}{144 \pi^2 \epsilon}.
\end{split}
\end{equation}
Instead of giving the vertex counterterms, with which the coupling constant ones could be found through Eq.~(\ref{eq:counterterms}), we give the solution for the latter directly:

\begin{equation}
\begin{split}
\delta Z^\mathscr{H}_{g_{3L}} &= -\frac{3g_{3L}^2(53 \mathcal{O}_{12}^2 + \mathcal{O}_{22}^2 + 25) + 4\sqrt{3}g_{3L}g_{X}\mathcal{O}_{22}\mathcal{O}_{32}+4g_X^2 \mathcal{O}_{32}^2}{576 \pi^2 \epsilon} \\
\delta Z^\mathscr{H}_{g_{X}} &= \frac{\frac{28}{3}g_X^3 F_1+ \frac{2}{3}g_{3L}g_X^2 F_2 - g_{3L}^2 g_X F_3 - g_{3L}^3 F_4}{192 \pi^2 g_X \mathcal{O}_{33} \epsilon},
\end{split}
\end{equation}
where, 

\begin{equation}
\begin{split}
F_1 &= \mathcal{O}_{33} (\mathcal{O}_{32}^2 + 9\mathcal{O}_{33}^2) \\
F_2 &= 2 \mathcal{O}_{32}\mathcal{O}_{33}(3\mathcal{O}_{12}-\sqrt{3}\mathcal{O}_{22})+3\mathcal{O}_{13}(2\mathcal{O}_{32}^2+ 63 \mathcal{O}_{33}^2)-\sqrt{3}\mathcal{O}_{23}(2\mathcal{O}_{32}^2+ 79 \mathcal{O}_{33}^2) \\
F_3 &= \mathcal{O}_{12}\left[ 6\mathcal{O}_{13}\mathcal{O}_{32}-2\sqrt{3}(\mathcal{O}_{22}\mathcal{O}_{33}+\mathcal{O}_{23}\mathcal{O}_{32})\right]  -4 \sqrt{3} \mathcal{O}_{13}(\mathcal{O}_{22}\mathcal{O}_{32}-4\mathcal{O}_{23}\mathcal{O}_{33}) \\
& \quad + \mathcal{O}_{33}\left[ \mathcal{O}_{22}^2+2\mathcal{O}_{23}^2+10\mathcal{O}_{13}^2 +3 (\mathcal{O}_{12}^2+1) \right] + 4 \mathcal{O}_{22} \mathcal{O}_{23} \mathcal{O}_{32} \\
F_4 &= -3 \mathcal{O}_{13}(-5 \mathcal{O}_{13} + 28\mathcal{O}_{12}^2 - \sqrt{3} \mathcal{O}_{12} \mathcal{O}_{22} + \mathcal{O}_{22}^2 - 9\mathcal{O}_{23}^2 + 23)  \\
& \quad + \mathcal{O}_{23}\left[-3\mathcal{O}_{12}\mathcal{O}_{22}+\sqrt{3}(28\mathcal{O}_{12}^2+\mathcal{O}_{22}^2 - 9\mathcal{O}_{23}^2-5\mathcal{O}_{13}^2+ 35)\right].
\end{split}
\end{equation}

With these, the $\beta$-functions may be calculated through

\begin{equation}
\begin{split}
\beta_{g_X} &= g_{X}^2 \frac{\partial \delta Z^{(1)}_{g_X}}{\partial g_{X}}+ g_{X}g_{3L} \frac{\partial \delta Z^{(1)}_{g_{3L}}}{\partial g_{X}} \\
\beta_{g_{3L}} &= g_{3L}^2 \frac{\partial \delta Z^{(1)}_{g_{3L}}}{\partial g_{{3L}}}+ g_{3L}g_{X} \frac{\partial \delta Z^{(1)}_{g_{X}}}{\partial g_{{3L}}},
\end{split}
\end{equation}
where the $(1)$ superscript indicates the coefficient of the first order $1/\epsilon$ pole.

\afterpage{\clearpage}
\begin{figure*}[p]
	\begin{center}	
 	\begin{adjustbox}{minipage=0.93\linewidth}
 		\begin{center}
		\includegraphics[width=0.33\linewidth]{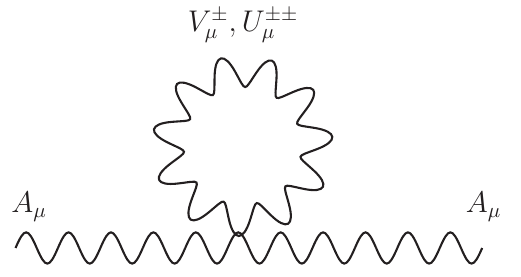}
		\includegraphics[width=0.33\linewidth]{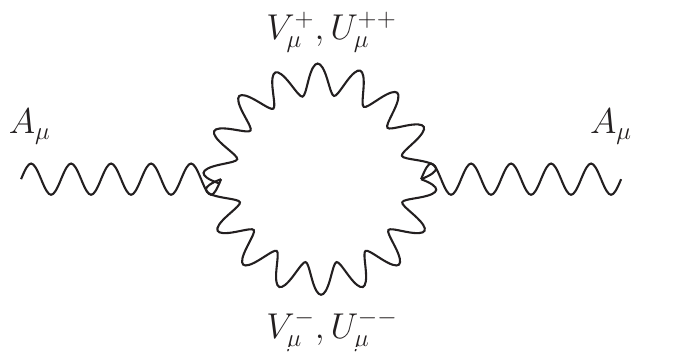} \\
		\vspace*{4mm}
		\includegraphics[width=0.33\linewidth]{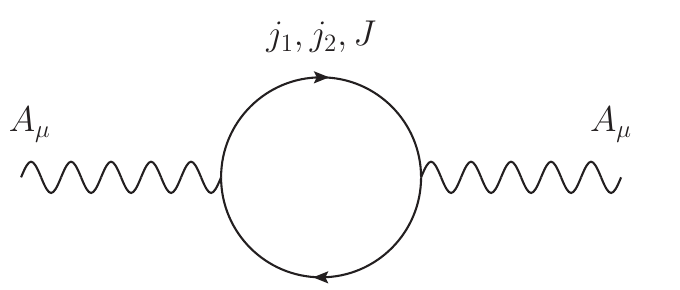}
		\includegraphics[width=0.33\linewidth]{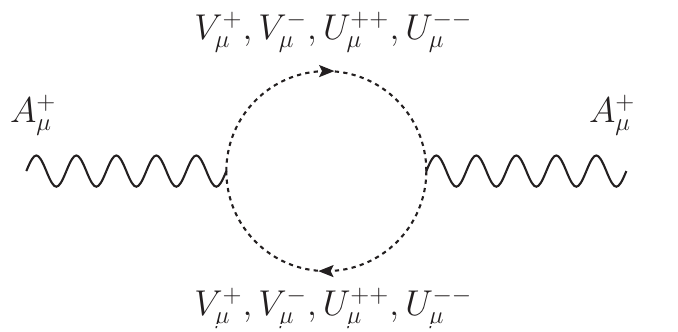} \\
		\vspace*{4mm}
		\includegraphics[width=0.32\linewidth]{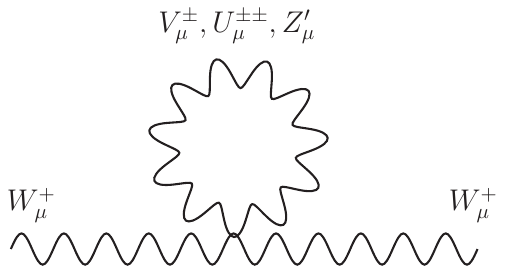}
		\includegraphics[width=0.32\linewidth]{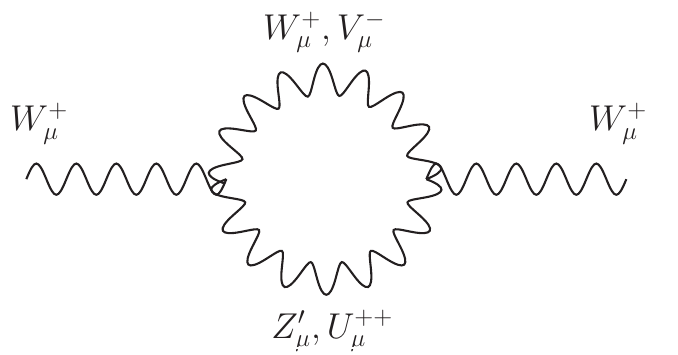}
		\includegraphics[width=0.32\linewidth]{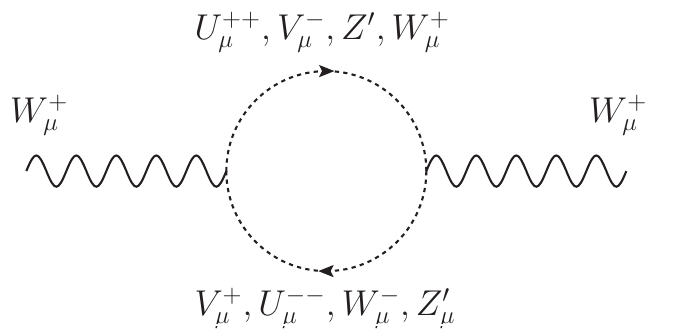} \\
		\vspace*{4mm}
		\includegraphics[width=0.33\linewidth]{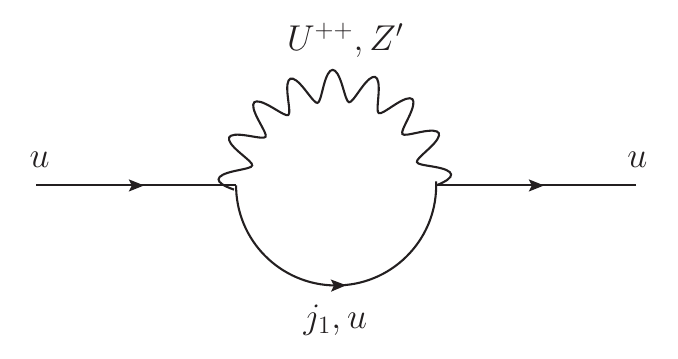}
		\includegraphics[width=0.33\linewidth]{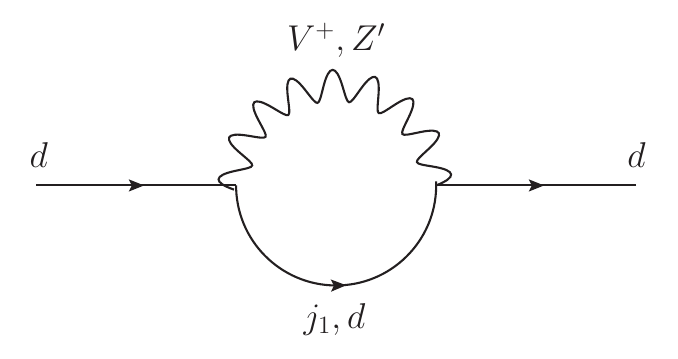} \\
		\vspace*{4mm}
		\includegraphics[width=0.33\linewidth]{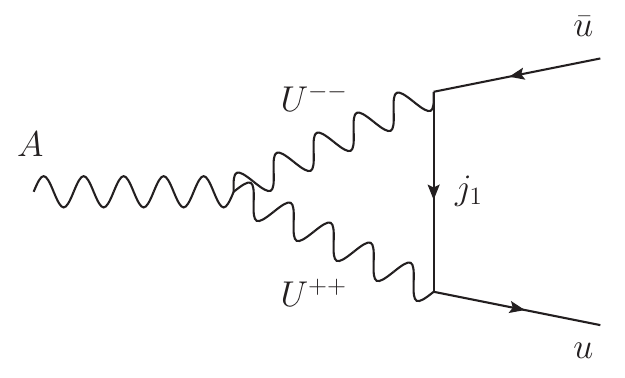}
		\includegraphics[width=0.33\linewidth]{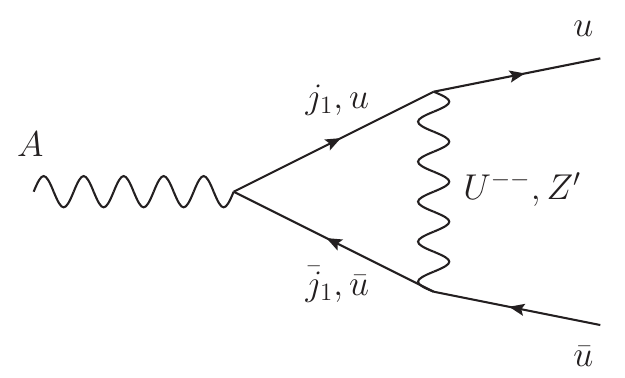}\\
		\vspace*{4mm}
		\includegraphics[width=0.33\linewidth]{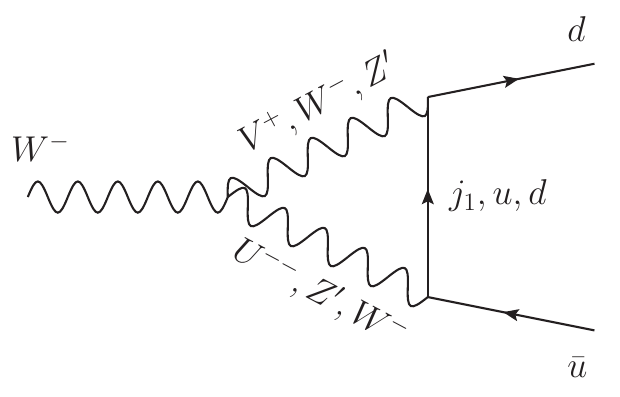}
		\includegraphics[width=0.33\linewidth]{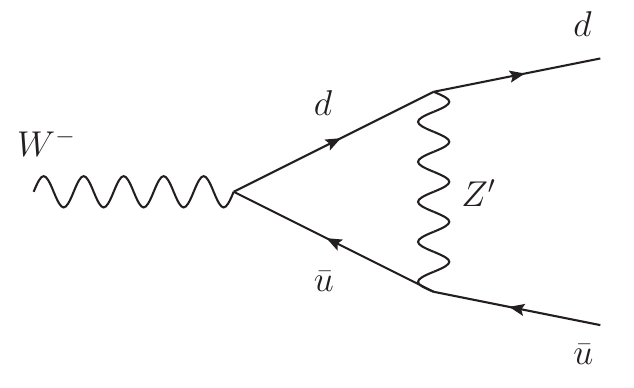}
		\end{center}
 	\end{adjustbox}
	\end{center}
	\caption{The complete set of exotic diagrams of the m331 relevant to the RGE analysis of the $\gamma u \bar{u}$ and $W d\bar{u}$ vertices. Time flows from left to right. To avoid introducing new symbols and since they only appear here, ghosts are denoted by their corresponding vector boson label, and are depicted by dotted lines. Diagrams with $N$ labels on internal lines should be read as $N$ diagrams, each with a combination of labels that should be paired as they are ordered. To allow for a sanity check, we note that the number of diagrams counts 31.}
	\label{fig:diagrams}
\end{figure*}

\clearpage

\bibliography{biblio}

\begin{thebibliography}{50}%
\makeatletter
\providecommand \@ifxundefined [1]{%
 \@ifx{#1\undefined}
}%
\providecommand \@ifnum [1]{%
 \ifnum #1\expandafter \@firstoftwo
 \else \expandafter \@secondoftwo
 \fi
}%
\providecommand \@ifx [1]{%
 \ifx #1\expandafter \@firstoftwo
 \else \expandafter \@secondoftwo
 \fi
}%
\providecommand \natexlab [1]{#1}%
\providecommand \enquote  [1]{``#1''}%
\providecommand \bibnamefont  [1]{#1}%
\providecommand \bibfnamefont [1]{#1}%
\providecommand \citenamefont [1]{#1}%
\providecommand \href@noop [0]{\@secondoftwo}%
\providecommand \href [0]{\begingroup \@sanitize@url \@href}%
\providecommand \@href[1]{\@@startlink{#1}\@@href}%
\providecommand \@@href[1]{\endgroup#1\@@endlink}%
\providecommand \@sanitize@url [0]{\catcode `\\12\catcode `\$12\catcode
  `\&12\catcode `\#12\catcode `\^12\catcode `\_12\catcode `\%12\relax}%
\providecommand \@@startlink[1]{}%
\providecommand \@@endlink[0]{}%
\providecommand \url  [0]{\begingroup\@sanitize@url \@url }%
\providecommand \@url [1]{\endgroup\@href {#1}{\urlprefix }}%
\providecommand \urlprefix  [0]{URL }%
\providecommand \Eprint [0]{\href }%
\providecommand \doibase [0]{https://doi.org/}%
\providecommand \selectlanguage [0]{\@gobble}%
\providecommand \bibinfo  [0]{\@secondoftwo}%
\providecommand \bibfield  [0]{\@secondoftwo}%
\providecommand \translation [1]{[#1]}%
\providecommand \BibitemOpen [0]{}%
\providecommand \bibitemStop [0]{}%
\providecommand \bibitemNoStop [0]{.\EOS\space}%
\providecommand \EOS [0]{\spacefactor3000\relax}%
\providecommand \BibitemShut  [1]{\csname bibitem#1\endcsname}%
\let\auto@bib@innerbib\@empty
\bibitem [{\citenamefont {Martin}(1998)}]{Martin:1997ns}%
  \BibitemOpen
  \bibfield  {author} {\bibinfo {author} {\bibfnamefont {S.~P.}\ \bibnamefont
  {Martin}},\ }\bibfield  {title} {\bibinfo {title} {{A Supersymmetry
  primer}},\ }\href {https://doi.org/10.1142/9789812839657_0001} {\bibfield
  {journal} {\bibinfo  {journal} {Adv. Ser. Direct. High Energy Phys.}\
  }\textbf {\bibinfo {volume} {18}},\ \bibinfo {pages} {1} (\bibinfo {year}
  {1998})},\ \Eprint {https://arxiv.org/abs/hep-ph/9709356}
  {arXiv:hep-ph/9709356} \BibitemShut {NoStop}%
\bibitem [{\citenamefont {Buras}\ \emph {et~al.}(1978)\citenamefont {Buras},
  \citenamefont {Ellis}, \citenamefont {Gaillard},\ and\ \citenamefont
  {Nanopoulos}}]{Buras:1977yy}%
  \BibitemOpen
  \bibfield  {author} {\bibinfo {author} {\bibfnamefont {A.~J.}\ \bibnamefont
  {Buras}}, \bibinfo {author} {\bibfnamefont {J.~R.}\ \bibnamefont {Ellis}},
  \bibinfo {author} {\bibfnamefont {M.~K.}\ \bibnamefont {Gaillard}},\ and\
  \bibinfo {author} {\bibfnamefont {D.~V.}\ \bibnamefont {Nanopoulos}},\
  }\bibfield  {title} {\bibinfo {title} {{Aspects of the Grand Unification of
  Strong, Weak and Electromagnetic Interactions}},\ }\href
  {https://doi.org/10.1016/0550-3213(78)90214-6} {\bibfield  {journal}
  {\bibinfo  {journal} {Nucl. Phys. B}\ }\textbf {\bibinfo {volume} {135}},\
  \bibinfo {pages} {66} (\bibinfo {year} {1978})}\BibitemShut {NoStop}%
\bibitem [{\citenamefont {Langacker}(1981)}]{Langacker:1980js}%
  \BibitemOpen
  \bibfield  {author} {\bibinfo {author} {\bibfnamefont {P.}~\bibnamefont
  {Langacker}},\ }\bibfield  {title} {\bibinfo {title} {{Grand Unified Theories
  and Proton Decay}},\ }\href {https://doi.org/10.1016/0370-1573(81)90059-4}
  {\bibfield  {journal} {\bibinfo  {journal} {Phys. Rept.}\ }\textbf {\bibinfo
  {volume} {72}},\ \bibinfo {pages} {185} (\bibinfo {year} {1981})}\BibitemShut
  {NoStop}%
\bibitem [{\citenamefont {Aad}\ \emph {et~al.}(2020)\citenamefont {Aad} \emph
  {et~al.}}]{ATLAS:2019lng}%
  \BibitemOpen
  \bibfield  {author} {\bibinfo {author} {\bibfnamefont {G.}~\bibnamefont
  {Aad}} \emph {et~al.} (\bibinfo {collaboration} {ATLAS}),\ }\bibfield
  {title} {\bibinfo {title} {{Searches for electroweak production of
  supersymmetric particles with compressed mass spectra in $\sqrt{s}=$ 13 TeV
  $pp$ collisions with the ATLAS detector}},\ }\href
  {https://doi.org/10.1103/PhysRevD.101.052005} {\bibfield  {journal} {\bibinfo
   {journal} {Phys. Rev. D}\ }\textbf {\bibinfo {volume} {101}},\ \bibinfo
  {pages} {052005} (\bibinfo {year} {2020})},\ \Eprint
  {https://arxiv.org/abs/1911.12606} {arXiv:1911.12606 [hep-ex]} \BibitemShut
  {NoStop}%
\bibitem [{\citenamefont {Collaboration}\ \emph {et~al.}(2019)\citenamefont
  {Collaboration} \emph {et~al.}}]{CMS:2019zmd}%
  \BibitemOpen
  \bibfield  {author} {\bibinfo {author} {\bibfnamefont {T.~C.}\ \bibnamefont
  {Collaboration}} \emph {et~al.} (\bibinfo {collaboration} {CMS}),\ }\bibfield
   {title} {\bibinfo {title} {{Search for supersymmetry in proton-proton
  collisions at 13 TeV in final states with jets and missing transverse
  momentum}},\ }\href {https://doi.org/10.1007/JHEP10(2019)244} {\bibfield
  {journal} {\bibinfo  {journal} {JHEP}\ }\textbf {\bibinfo {volume} {10}},\
  \bibinfo {pages} {244}},\ \Eprint {https://arxiv.org/abs/1908.04722}
  {arXiv:1908.04722 [hep-ex]} \BibitemShut {NoStop}%
\bibitem [{\citenamefont {Feldman}\ \emph {et~al.}(2007)\citenamefont
  {Feldman}, \citenamefont {Liu},\ and\ \citenamefont {Nath}}]{Feldman:2007wj}%
  \BibitemOpen
  \bibfield  {author} {\bibinfo {author} {\bibfnamefont {D.}~\bibnamefont
  {Feldman}}, \bibinfo {author} {\bibfnamefont {Z.}~\bibnamefont {Liu}},\ and\
  \bibinfo {author} {\bibfnamefont {P.}~\bibnamefont {Nath}},\ }\bibfield
  {title} {\bibinfo {title} {{The Stueckelberg Z-prime Extension with Kinetic
  Mixing and Milli-Charged Dark Matter From the Hidden Sector}},\ }\href
  {https://doi.org/10.1103/PhysRevD.75.115001} {\bibfield  {journal} {\bibinfo
  {journal} {Phys. Rev. D}\ }\textbf {\bibinfo {volume} {75}},\ \bibinfo
  {pages} {115001} (\bibinfo {year} {2007})},\ \Eprint
  {https://arxiv.org/abs/hep-ph/0702123} {arXiv:hep-ph/0702123} \BibitemShut
  {NoStop}%
\bibitem [{\citenamefont {Kors}\ and\ \citenamefont
  {Nath}(2004)}]{Kors:2004dx}%
  \BibitemOpen
  \bibfield  {author} {\bibinfo {author} {\bibfnamefont {B.}~\bibnamefont
  {Kors}}\ and\ \bibinfo {author} {\bibfnamefont {P.}~\bibnamefont {Nath}},\
  }\bibfield  {title} {\bibinfo {title} {{A Stueckelberg extension of the
  standard model}},\ }\href {https://doi.org/10.1016/j.physletb.2004.02.051}
  {\bibfield  {journal} {\bibinfo  {journal} {Phys. Lett. B}\ }\textbf
  {\bibinfo {volume} {586}},\ \bibinfo {pages} {366} (\bibinfo {year}
  {2004})},\ \Eprint {https://arxiv.org/abs/hep-ph/0402047}
  {arXiv:hep-ph/0402047} \BibitemShut {NoStop}%
\bibitem [{\citenamefont {He}\ \emph {et~al.}(1991)\citenamefont {He},
  \citenamefont {Joshi}, \citenamefont {Lew},\ and\ \citenamefont
  {Volkas}}]{He:1991qd}%
  \BibitemOpen
  \bibfield  {author} {\bibinfo {author} {\bibfnamefont {X.-G.}\ \bibnamefont
  {He}}, \bibinfo {author} {\bibfnamefont {G.~C.}\ \bibnamefont {Joshi}},
  \bibinfo {author} {\bibfnamefont {H.}~\bibnamefont {Lew}},\ and\ \bibinfo
  {author} {\bibfnamefont {R.~R.}\ \bibnamefont {Volkas}},\ }\bibfield  {title}
  {\bibinfo {title} {{Simplest Z-prime model}},\ }\href
  {https://doi.org/10.1103/PhysRevD.44.2118} {\bibfield  {journal} {\bibinfo
  {journal} {Phys. Rev. D}\ }\textbf {\bibinfo {volume} {44}},\ \bibinfo
  {pages} {2118} (\bibinfo {year} {1991})}\BibitemShut {NoStop}%
\bibitem [{\citenamefont {Pisano}\ and\ \citenamefont
  {Pleitez}(1995)}]{Pisano:1994tf}%
  \BibitemOpen
  \bibfield  {author} {\bibinfo {author} {\bibfnamefont {F.}~\bibnamefont
  {Pisano}}\ and\ \bibinfo {author} {\bibfnamefont {V.}~\bibnamefont
  {Pleitez}},\ }\bibfield  {title} {\bibinfo {title} {{SU(4)-L x U(1)-N model
  for the electroweak interactions}},\ }\href
  {https://doi.org/10.1103/PhysRevD.51.3865} {\bibfield  {journal} {\bibinfo
  {journal} {Phys. Rev. D}\ }\textbf {\bibinfo {volume} {51}},\ \bibinfo
  {pages} {3865} (\bibinfo {year} {1995})},\ \Eprint
  {https://arxiv.org/abs/hep-ph/9401272} {arXiv:hep-ph/9401272} \BibitemShut
  {NoStop}%
\bibitem [{\citenamefont {Senjanovic}\ and\ \citenamefont
  {Mohapatra}(1975)}]{Senjanovic:1975rk}%
  \BibitemOpen
  \bibfield  {author} {\bibinfo {author} {\bibfnamefont {G.}~\bibnamefont
  {Senjanovic}}\ and\ \bibinfo {author} {\bibfnamefont {R.~N.}\ \bibnamefont
  {Mohapatra}},\ }\bibfield  {title} {\bibinfo {title} {{Exact Left-Right
  Symmetry and Spontaneous Violation of Parity}},\ }\href
  {https://doi.org/10.1103/PhysRevD.12.1502} {\bibfield  {journal} {\bibinfo
  {journal} {Phys. Rev. D}\ }\textbf {\bibinfo {volume} {12}},\ \bibinfo
  {pages} {1502} (\bibinfo {year} {1975})}\BibitemShut {NoStop}%
\bibitem [{\citenamefont {Mohapatra}\ and\ \citenamefont
  {Pati}(1975)}]{Mohapatra:1974hk}%
  \BibitemOpen
  \bibfield  {author} {\bibinfo {author} {\bibfnamefont {R.~N.}\ \bibnamefont
  {Mohapatra}}\ and\ \bibinfo {author} {\bibfnamefont {J.~C.}\ \bibnamefont
  {Pati}},\ }\bibfield  {title} {\bibinfo {title} {{Left-Right Gauge Symmetry
  and an Isoconjugate Model of CP Violation}},\ }\href
  {https://doi.org/10.1103/PhysRevD.11.566} {\bibfield  {journal} {\bibinfo
  {journal} {Phys. Rev. D}\ }\textbf {\bibinfo {volume} {11}},\ \bibinfo
  {pages} {566} (\bibinfo {year} {1975})}\BibitemShut {NoStop}%
\bibitem [{\citenamefont {Deshpande}\ \emph {et~al.}(1991)\citenamefont
  {Deshpande}, \citenamefont {Gunion}, \citenamefont {Kayser},\ and\
  \citenamefont {Olness}}]{Deshpande:1990ip}%
  \BibitemOpen
  \bibfield  {author} {\bibinfo {author} {\bibfnamefont {N.~G.}\ \bibnamefont
  {Deshpande}}, \bibinfo {author} {\bibfnamefont {J.~F.}\ \bibnamefont
  {Gunion}}, \bibinfo {author} {\bibfnamefont {B.}~\bibnamefont {Kayser}},\
  and\ \bibinfo {author} {\bibfnamefont {F.~I.}\ \bibnamefont {Olness}},\
  }\bibfield  {title} {\bibinfo {title} {{Left-right symmetric electroweak
  models with triplet Higgs}},\ }\href
  {https://doi.org/10.1103/PhysRevD.44.837} {\bibfield  {journal} {\bibinfo
  {journal} {Phys. Rev. D}\ }\textbf {\bibinfo {volume} {44}},\ \bibinfo
  {pages} {837} (\bibinfo {year} {1991})}\BibitemShut {NoStop}%
\bibitem [{\citenamefont {Pisano}\ and\ \citenamefont
  {Pleitez}(1992)}]{Pisano:1992bxx}%
  \BibitemOpen
  \bibfield  {author} {\bibinfo {author} {\bibfnamefont {F.}~\bibnamefont
  {Pisano}}\ and\ \bibinfo {author} {\bibfnamefont {V.}~\bibnamefont
  {Pleitez}},\ }\bibfield  {title} {\bibinfo {title} {{An SU(3) x U(1) model
  for electroweak interactions}},\ }\href
  {https://doi.org/10.1103/PhysRevD.46.410} {\bibfield  {journal} {\bibinfo
  {journal} {Phys. Rev. D}\ }\textbf {\bibinfo {volume} {46}},\ \bibinfo
  {pages} {410} (\bibinfo {year} {1992})},\ \Eprint
  {https://arxiv.org/abs/hep-ph/9206242} {arXiv:hep-ph/9206242} \BibitemShut
  {NoStop}%
\bibitem [{\citenamefont {Frampton}(1992)}]{Frampton:1992wt}%
  \BibitemOpen
  \bibfield  {author} {\bibinfo {author} {\bibfnamefont {P.~H.}\ \bibnamefont
  {Frampton}},\ }\bibfield  {title} {\bibinfo {title} {{Chiral dilepton model
  and the flavor question}},\ }\href
  {https://doi.org/10.1103/PhysRevLett.69.2889} {\bibfield  {journal} {\bibinfo
   {journal} {Phys. Rev. Lett.}\ }\textbf {\bibinfo {volume} {69}},\ \bibinfo
  {pages} {2889} (\bibinfo {year} {1992})}\BibitemShut {NoStop}%
\bibitem [{\citenamefont {Montero}\ \emph {et~al.}(1993)\citenamefont
  {Montero}, \citenamefont {Pisano},\ and\ \citenamefont
  {Pleitez}}]{PhysRevD.47.2918}%
  \BibitemOpen
  \bibfield  {author} {\bibinfo {author} {\bibfnamefont {J.~C.}\ \bibnamefont
  {Montero}}, \bibinfo {author} {\bibfnamefont {F.}~\bibnamefont {Pisano}},\
  and\ \bibinfo {author} {\bibfnamefont {V.}~\bibnamefont {Pleitez}},\
  }\bibfield  {title} {\bibinfo {title} {Neutral currents and
  glashow-iliopoulos-maiani mechanism in
  $\mathrm{SU}{(3)}_{L}\ensuremath{\bigotimes}\mathrm{U}{(1)}_{N}$ models for
  electroweak interactions},\ }\href {https://doi.org/10.1103/PhysRevD.47.2918}
  {\bibfield  {journal} {\bibinfo  {journal} {Phys. Rev. D}\ }\textbf {\bibinfo
  {volume} {47}},\ \bibinfo {pages} {2918} (\bibinfo {year}
  {1993})}\BibitemShut {NoStop}%
\bibitem [{\citenamefont {Foot}\ \emph {et~al.}(1994)\citenamefont {Foot},
  \citenamefont {Long},\ and\ \citenamefont {Tran}}]{PhysRevD.50.R34}%
  \BibitemOpen
  \bibfield  {author} {\bibinfo {author} {\bibfnamefont {R.}~\bibnamefont
  {Foot}}, \bibinfo {author} {\bibfnamefont {H.~N.}\ \bibnamefont {Long}},\
  and\ \bibinfo {author} {\bibfnamefont {T.~A.}\ \bibnamefont {Tran}},\
  }\bibfield  {title} {\bibinfo {title}
  {Su(3${)}_{\mathit{l}}$\ensuremath{\bigotimes}u(1${)}_{\mathit{n}}$ and
  su(4${)}_{\mathit{l}}$\ensuremath{\bigotimes}u(1${)}_{\mathit{n}}$ gauge
  models with right-handed neutrinos},\ }\href
  {https://doi.org/10.1103/PhysRevD.50.R34} {\bibfield  {journal} {\bibinfo
  {journal} {Phys. Rev. D}\ }\textbf {\bibinfo {volume} {50}},\ \bibinfo
  {pages} {R34} (\bibinfo {year} {1994})}\BibitemShut {NoStop}%
\bibitem [{\citenamefont {Long}(1996)}]{PhysRevD.53.437}%
  \BibitemOpen
  \bibfield  {author} {\bibinfo {author} {\bibfnamefont {H.~N.}\ \bibnamefont
  {Long}},\ }\bibfield  {title} {\bibinfo {title}
  {Su(3${)}_{\mathit{c}}$\ensuremath{\bigotimes}su(3${)}_{\mathit{l}}$\ensuremath{\bigotimes}u(1${)}_{\mathit{n}}$
  model with right-handed neutrinos},\ }\href
  {https://doi.org/10.1103/PhysRevD.53.437} {\bibfield  {journal} {\bibinfo
  {journal} {Phys. Rev. D}\ }\textbf {\bibinfo {volume} {53}},\ \bibinfo
  {pages} {437} (\bibinfo {year} {1996})}\BibitemShut {NoStop}%
\bibitem [{\citenamefont {Pleitez}\ and\ \citenamefont
  {Tonasse}(1993)}]{PhysRevD.48.2353}%
  \BibitemOpen
  \bibfield  {author} {\bibinfo {author} {\bibfnamefont {V.}~\bibnamefont
  {Pleitez}}\ and\ \bibinfo {author} {\bibfnamefont {M.~D.}\ \bibnamefont
  {Tonasse}},\ }\bibfield  {title} {\bibinfo {title} {Heavy charged leptons in
  an $\mathrm{SU}{(3)}_{L}\ensuremath{\bigotimes}\mathrm{U}{(1)}_{N}$ model},\
  }\href {https://doi.org/10.1103/PhysRevD.48.2353} {\bibfield  {journal}
  {\bibinfo  {journal} {Phys. Rev. D}\ }\textbf {\bibinfo {volume} {48}},\
  \bibinfo {pages} {2353} (\bibinfo {year} {1993})}\BibitemShut {NoStop}%
\bibitem [{\citenamefont {Dong}\ \emph {et~al.}(2014)\citenamefont {Dong},
  \citenamefont {Ngan},\ and\ \citenamefont {Soa}}]{Dong:2014esa}%
  \BibitemOpen
  \bibfield  {author} {\bibinfo {author} {\bibfnamefont {P.~V.}\ \bibnamefont
  {Dong}}, \bibinfo {author} {\bibfnamefont {N.~T.~K.}\ \bibnamefont {Ngan}},\
  and\ \bibinfo {author} {\bibfnamefont {D.~V.}\ \bibnamefont {Soa}},\
  }\bibfield  {title} {\bibinfo {title} {{Simple 3-3-1 model and implication
  for dark matter}},\ }\href {https://doi.org/10.1103/PhysRevD.90.075019}
  {\bibfield  {journal} {\bibinfo  {journal} {Phys. Rev. D}\ }\textbf {\bibinfo
  {volume} {90}},\ \bibinfo {pages} {075019} (\bibinfo {year} {2014})},\
  \Eprint {https://arxiv.org/abs/1407.3839} {arXiv:1407.3839 [hep-ph]}
  \BibitemShut {NoStop}%
\bibitem [{\citenamefont {Dong}\ \emph {et~al.}(2015)\citenamefont {Dong},
  \citenamefont {Kim}, \citenamefont {Soa},\ and\ \citenamefont
  {Thuy}}]{Dong:2015rka}%
  \BibitemOpen
  \bibfield  {author} {\bibinfo {author} {\bibfnamefont {P.~V.}\ \bibnamefont
  {Dong}}, \bibinfo {author} {\bibfnamefont {C.~S.}\ \bibnamefont {Kim}},
  \bibinfo {author} {\bibfnamefont {D.~V.}\ \bibnamefont {Soa}},\ and\ \bibinfo
  {author} {\bibfnamefont {N.~T.}\ \bibnamefont {Thuy}},\ }\bibfield  {title}
  {\bibinfo {title} {{Investigation of Dark Matter in Minimal 3-3-1 Models}},\
  }\href {https://doi.org/10.1103/PhysRevD.91.115019} {\bibfield  {journal}
  {\bibinfo  {journal} {Phys. Rev. D}\ }\textbf {\bibinfo {volume} {91}},\
  \bibinfo {pages} {115019} (\bibinfo {year} {2015})},\ \Eprint
  {https://arxiv.org/abs/1501.04385} {arXiv:1501.04385 [hep-ph]} \BibitemShut
  {NoStop}%
\bibitem [{\citenamefont {Huong}\ \emph {et~al.}(2019)\citenamefont {Huong},
  \citenamefont {Dinh}, \citenamefont {Thien},\ and\ \citenamefont
  {Van~Dong}}]{Huong:2019vej}%
  \BibitemOpen
  \bibfield  {author} {\bibinfo {author} {\bibfnamefont {D.~T.}\ \bibnamefont
  {Huong}}, \bibinfo {author} {\bibfnamefont {D.~N.}\ \bibnamefont {Dinh}},
  \bibinfo {author} {\bibfnamefont {L.~D.}\ \bibnamefont {Thien}},\ and\
  \bibinfo {author} {\bibfnamefont {P.}~\bibnamefont {Van~Dong}},\ }\bibfield
  {title} {\bibinfo {title} {{Dark matter and flavor changing in the flipped
  3-3-1 model}},\ }\href {https://doi.org/10.1007/JHEP08(2019)051} {\bibfield
  {journal} {\bibinfo  {journal} {JHEP}\ }\textbf {\bibinfo {volume} {08}},\
  \bibinfo {pages} {051}},\ \Eprint {https://arxiv.org/abs/1906.05240}
  {arXiv:1906.05240 [hep-ph]} \BibitemShut {NoStop}%
\bibitem [{\citenamefont {Deppisch}\ \emph {et~al.}(2016)\citenamefont
  {Deppisch}, \citenamefont {Hati}, \citenamefont {Patra}, \citenamefont
  {Sarkar},\ and\ \citenamefont {Valle}}]{Deppisch:2016jzl}%
  \BibitemOpen
  \bibfield  {author} {\bibinfo {author} {\bibfnamefont {F.~F.}\ \bibnamefont
  {Deppisch}}, \bibinfo {author} {\bibfnamefont {C.}~\bibnamefont {Hati}},
  \bibinfo {author} {\bibfnamefont {S.}~\bibnamefont {Patra}}, \bibinfo
  {author} {\bibfnamefont {U.}~\bibnamefont {Sarkar}},\ and\ \bibinfo {author}
  {\bibfnamefont {J.~W.~F.}\ \bibnamefont {Valle}},\ }\bibfield  {title}
  {\bibinfo {title} {{331 Models and Grand Unification: From Minimal SU(5) to
  Minimal SU(6)}},\ }\href {https://doi.org/10.1016/j.physletb.2016.10.002}
  {\bibfield  {journal} {\bibinfo  {journal} {Phys. Lett. B}\ }\textbf
  {\bibinfo {volume} {762}},\ \bibinfo {pages} {432} (\bibinfo {year}
  {2016})},\ \Eprint {https://arxiv.org/abs/1608.05334} {arXiv:1608.05334
  [hep-ph]} \BibitemShut {NoStop}%
\bibitem [{\citenamefont {Diaz}\ \emph {et~al.}(2007)\citenamefont {Diaz},
  \citenamefont {Gallego},\ and\ \citenamefont {Martinez}}]{Diaz:2005bw}%
  \BibitemOpen
  \bibfield  {author} {\bibinfo {author} {\bibfnamefont {R.~A.}\ \bibnamefont
  {Diaz}}, \bibinfo {author} {\bibfnamefont {D.}~\bibnamefont {Gallego}},\ and\
  \bibinfo {author} {\bibfnamefont {R.}~\bibnamefont {Martinez}},\ }\bibfield
  {title} {\bibinfo {title} {{Renormalization group and grand unification with
  331 models}},\ }\href {https://doi.org/10.1142/S0217751X07036142} {\bibfield
  {journal} {\bibinfo  {journal} {Int. J. Mod. Phys. A}\ }\textbf {\bibinfo
  {volume} {22}},\ \bibinfo {pages} {1849} (\bibinfo {year} {2007})},\ \Eprint
  {https://arxiv.org/abs/hep-ph/0505096} {arXiv:hep-ph/0505096} \BibitemShut
  {NoStop}%
\bibitem [{\citenamefont {Frampton}\ and\ \citenamefont
  {Lee}(1990)}]{PhysRevLett.64.619}%
  \BibitemOpen
  \bibfield  {author} {\bibinfo {author} {\bibfnamefont {P.~H.}\ \bibnamefont
  {Frampton}}\ and\ \bibinfo {author} {\bibfnamefont {B.-H.}\ \bibnamefont
  {Lee}},\ }\bibfield  {title} {\bibinfo {title} {Su(15) grand unification},\
  }\href {https://doi.org/10.1103/PhysRevLett.64.619} {\bibfield  {journal}
  {\bibinfo  {journal} {Phys. Rev. Lett.}\ }\textbf {\bibinfo {volume} {64}},\
  \bibinfo {pages} {619} (\bibinfo {year} {1990})}\BibitemShut {NoStop}%
\bibitem [{\citenamefont {Pal}(1992)}]{Pal:1991nfm}%
  \BibitemOpen
  \bibfield  {author} {\bibinfo {author} {\bibfnamefont {P.~B.}\ \bibnamefont
  {Pal}},\ }\bibfield  {title} {\bibinfo {title} {{Proton decay modes in SU(15)
  grand unification}},\ }\href {https://doi.org/10.1103/PhysRevD.45.2566}
  {\bibfield  {journal} {\bibinfo  {journal} {Phys. Rev. D}\ }\textbf {\bibinfo
  {volume} {45}},\ \bibinfo {pages} {2566} (\bibinfo {year}
  {1992})}\BibitemShut {NoStop}%
\bibitem [{\citenamefont {Dias}\ \emph {et~al.}(2005)\citenamefont {Dias},
  \citenamefont {Martinez},\ and\ \citenamefont {Pleitez}}]{Dias:2004dc}%
  \BibitemOpen
  \bibfield  {author} {\bibinfo {author} {\bibfnamefont {A.~G.}\ \bibnamefont
  {Dias}}, \bibinfo {author} {\bibfnamefont {R.}~\bibnamefont {Martinez}},\
  and\ \bibinfo {author} {\bibfnamefont {V.}~\bibnamefont {Pleitez}},\
  }\bibfield  {title} {\bibinfo {title} {{Concerning the Landau pole in 3-3-1
  models}},\ }\href {https://doi.org/10.1140/epjc/s2004-02083-0} {\bibfield
  {journal} {\bibinfo  {journal} {Eur. Phys. J. C}\ }\textbf {\bibinfo {volume}
  {39}},\ \bibinfo {pages} {101} (\bibinfo {year} {2005})},\ \Eprint
  {https://arxiv.org/abs/hep-ph/0407141} {arXiv:hep-ph/0407141} \BibitemShut
  {NoStop}%
\bibitem [{\citenamefont {Martinez}\ and\ \citenamefont
  {Ochoa}(2007)}]{Martinez:2006gb}%
  \BibitemOpen
  \bibfield  {author} {\bibinfo {author} {\bibfnamefont {R.}~\bibnamefont
  {Martinez}}\ and\ \bibinfo {author} {\bibfnamefont {F.}~\bibnamefont
  {Ochoa}},\ }\bibfield  {title} {\bibinfo {title} {{The Landau pole and
  Z-prime decays in the 331 bilepton model}},\ }\href
  {https://doi.org/10.1140/epjc/s10052-007-0307-6} {\bibfield  {journal}
  {\bibinfo  {journal} {Eur. Phys. J. C}\ }\textbf {\bibinfo {volume} {51}},\
  \bibinfo {pages} {701} (\bibinfo {year} {2007})},\ \Eprint
  {https://arxiv.org/abs/hep-ph/0606173} {arXiv:hep-ph/0606173} \BibitemShut
  {NoStop}%
\bibitem [{\citenamefont {Santos}\ and\ \citenamefont
  {Vasconcelos}(2018)}]{Santos:2017jbv}%
  \BibitemOpen
  \bibfield  {author} {\bibinfo {author} {\bibfnamefont {A.~C.~O.}\
  \bibnamefont {Santos}}\ and\ \bibinfo {author} {\bibfnamefont
  {P.}~\bibnamefont {Vasconcelos}},\ }\bibfield  {title} {\bibinfo {title}
  {{Lower Mass Bound on the $W^\prime$ mass via Neutrinoless Double Beta Decay
  in a 3-3-1 Model}},\ }\href {https://doi.org/10.1155/2018/9132381} {\bibfield
   {journal} {\bibinfo  {journal} {Adv. High Energy Phys.}\ }\textbf {\bibinfo
  {volume} {2018}},\ \bibinfo {pages} {9132381} (\bibinfo {year} {2018})},\
  \Eprint {https://arxiv.org/abs/1708.03955} {arXiv:1708.03955 [hep-ph]}
  \BibitemShut {NoStop}%
\bibitem [{\citenamefont {Doff}\ and\ \citenamefont
  {de~S.~Pires}(2023)}]{Doff:2023bgy}%
  \BibitemOpen
  \bibfield  {author} {\bibinfo {author} {\bibfnamefont {A.}~\bibnamefont
  {Doff}}\ and\ \bibinfo {author} {\bibfnamefont {C.~A.}\ \bibnamefont
  {de~S.~Pires}},\ }\bibfield  {title} {\bibinfo {title} {{Evading the Landau
  pole in the minimal 3-3-1 model with leptoquarks}},\ }\href
  {https://doi.org/10.1016/j.nuclphysb.2023.116254} {\bibfield  {journal}
  {\bibinfo  {journal} {Nucl. Phys. B}\ }\textbf {\bibinfo {volume} {992}},\
  \bibinfo {pages} {116254} (\bibinfo {year} {2023})},\ \Eprint
  {https://arxiv.org/abs/2302.08578} {arXiv:2302.08578 [hep-ph]} \BibitemShut
  {NoStop}%
\bibitem [{\citenamefont {Georgi}\ and\ \citenamefont
  {Weinberg}(1978)}]{Georgi:1977wk}%
  \BibitemOpen
  \bibfield  {author} {\bibinfo {author} {\bibfnamefont {H.}~\bibnamefont
  {Georgi}}\ and\ \bibinfo {author} {\bibfnamefont {S.}~\bibnamefont
  {Weinberg}},\ }\bibfield  {title} {\bibinfo {title} {{Neutral Currents in
  Expanded Gauge Theories}},\ }\href {https://doi.org/10.1103/PhysRevD.17.275}
  {\bibfield  {journal} {\bibinfo  {journal} {Phys. Rev. D}\ }\textbf {\bibinfo
  {volume} {17}},\ \bibinfo {pages} {275} (\bibinfo {year} {1978})}\BibitemShut
  {NoStop}%
\bibitem [{\citenamefont {Roy}(2019)}]{Roy:2019jqs}%
  \BibitemOpen
  \bibfield  {author} {\bibinfo {author} {\bibfnamefont {J.}~\bibnamefont
  {Roy}},\ }\bibfield  {title} {\bibinfo {title} {{Calculating $\beta$-function
  coefficients of Renormalization Group Equations}},\ }\href@noop {} {\
  (\bibinfo {year} {2019})},\ \Eprint {https://arxiv.org/abs/1907.10238}
  {arXiv:1907.10238 [hep-ph]} \BibitemShut {NoStop}%
\bibitem [{\citenamefont {Dias}\ \emph {et~al.}(2006)\citenamefont {Dias},
  \citenamefont {Montero},\ and\ \citenamefont {Pleitez}}]{Dias:2006ns}%
  \BibitemOpen
  \bibfield  {author} {\bibinfo {author} {\bibfnamefont {A.~G.}\ \bibnamefont
  {Dias}}, \bibinfo {author} {\bibfnamefont {J.~C.}\ \bibnamefont {Montero}},\
  and\ \bibinfo {author} {\bibfnamefont {V.}~\bibnamefont {Pleitez}},\
  }\bibfield  {title} {\bibinfo {title} {{Closing the SU(3)(L) x U(1)(X)
  symmetry at electroweak scale}},\ }\href
  {https://doi.org/10.1103/PhysRevD.73.113004} {\bibfield  {journal} {\bibinfo
  {journal} {Phys. Rev. D}\ }\textbf {\bibinfo {volume} {73}},\ \bibinfo
  {pages} {113004} (\bibinfo {year} {2006})},\ \Eprint
  {https://arxiv.org/abs/hep-ph/0605051} {arXiv:hep-ph/0605051} \BibitemShut
  {NoStop}%
\bibitem [{\citenamefont {Diaz}\ \emph {et~al.}(2004)\citenamefont {Diaz},
  \citenamefont {Mart\'{\i}nez},\ and\ \citenamefont
  {Ochoa}}]{PhysRevD.69.095009}%
  \BibitemOpen
  \bibfield  {author} {\bibinfo {author} {\bibfnamefont {R.~A.}\ \bibnamefont
  {Diaz}}, \bibinfo {author} {\bibfnamefont {R.}~\bibnamefont
  {Mart\'{\i}nez}},\ and\ \bibinfo {author} {\bibfnamefont {F.}~\bibnamefont
  {Ochoa}},\ }\bibfield  {title} {\bibinfo {title} {Scalar sector of the
  ${\mathrm{su}(3)}_{c}\ensuremath{\bigotimes}{\mathrm{su}(3)}_{L}\ensuremath{\bigotimes}{U(1)}_{X}$
  model},\ }\href {https://doi.org/10.1103/PhysRevD.69.095009} {\bibfield
  {journal} {\bibinfo  {journal} {Phys. Rev. D}\ }\textbf {\bibinfo {volume}
  {69}},\ \bibinfo {pages} {095009} (\bibinfo {year} {2004})}\BibitemShut
  {NoStop}%
\bibitem [{\citenamefont {Tully}\ and\ \citenamefont
  {Joshi}(2003)}]{Tully:1998wa}%
  \BibitemOpen
  \bibfield  {author} {\bibinfo {author} {\bibfnamefont {M.~B.}\ \bibnamefont
  {Tully}}\ and\ \bibinfo {author} {\bibfnamefont {G.~C.}\ \bibnamefont
  {Joshi}},\ }\bibfield  {title} {\bibinfo {title} {{The Scalar sector in 331
  models}},\ }\href {https://doi.org/10.1142/S0217751X03013995} {\bibfield
  {journal} {\bibinfo  {journal} {Int. J. Mod. Phys. A}\ }\textbf {\bibinfo
  {volume} {18}},\ \bibinfo {pages} {1573} (\bibinfo {year} {2003})},\ \Eprint
  {https://arxiv.org/abs/hep-ph/9810282} {arXiv:hep-ph/9810282} \BibitemShut
  {NoStop}%
\bibitem [{\citenamefont {Workman}\ \emph {et~al.}(2022)\citenamefont {Workman}
  \emph {et~al.}}]{ParticleDataGroup:2022pth}%
  \BibitemOpen
  \bibfield  {author} {\bibinfo {author} {\bibfnamefont {R.~L.}\ \bibnamefont
  {Workman}} \emph {et~al.} (\bibinfo {collaboration} {Particle Data Group}),\
  }\bibfield  {title} {\bibinfo {title} {{Review of Particle Physics}},\ }\href
  {https://doi.org/10.1093/ptep/ptac097} {\bibfield  {journal} {\bibinfo
  {journal} {PTEP}\ }\textbf {\bibinfo {volume} {2022}},\ \bibinfo {pages}
  {083C01} (\bibinfo {year} {2022})}\BibitemShut {NoStop}%
\bibitem [{\citenamefont {Appelquist}\ and\ \citenamefont
  {Carazzone}(1975)}]{Appelquist:1974tg}%
  \BibitemOpen
  \bibfield  {author} {\bibinfo {author} {\bibfnamefont {T.}~\bibnamefont
  {Appelquist}}\ and\ \bibinfo {author} {\bibfnamefont {J.}~\bibnamefont
  {Carazzone}},\ }\bibfield  {title} {\bibinfo {title} {{Infrared Singularities
  and Massive Fields}},\ }\href {https://doi.org/10.1103/PhysRevD.11.2856}
  {\bibfield  {journal} {\bibinfo  {journal} {Phys. Rev. D}\ }\textbf {\bibinfo
  {volume} {11}},\ \bibinfo {pages} {2856} (\bibinfo {year}
  {1975})}\BibitemShut {NoStop}%
\bibitem [{\citenamefont {Srednicki}(2007)}]{Srednicki:2007qs}%
  \BibitemOpen
  \bibfield  {author} {\bibinfo {author} {\bibfnamefont {M.}~\bibnamefont
  {Srednicki}},\ }\href@noop {} {\emph {\bibinfo {title} {{Quantum field
  theory}}}}\ (\bibinfo  {publisher} {Cambridge University Press},\ \bibinfo
  {year} {2007})\BibitemShut {NoStop}%
\bibitem [{\citenamefont {Weinberg}(2013)}]{Weinberg:1996kr}%
  \BibitemOpen
  \bibfield  {author} {\bibinfo {author} {\bibfnamefont {S.}~\bibnamefont
  {Weinberg}},\ }\href {https://doi.org/10.1017/CBO9781139644174} {\emph
  {\bibinfo {title} {{The quantum theory of fields. Vol. 2: Modern
  applications}}}}\ (\bibinfo  {publisher} {Cambridge University Press},\
  \bibinfo {year} {2013})\BibitemShut {NoStop}%
\bibitem [{\citenamefont {Denner}\ \emph {et~al.}(1992)\citenamefont {Denner},
  \citenamefont {Eck}, \citenamefont {Hahn},\ and\ \citenamefont
  {Kublbeck}}]{Denner:1992vza}%
  \BibitemOpen
  \bibfield  {author} {\bibinfo {author} {\bibfnamefont {A.}~\bibnamefont
  {Denner}}, \bibinfo {author} {\bibfnamefont {H.}~\bibnamefont {Eck}},
  \bibinfo {author} {\bibfnamefont {O.}~\bibnamefont {Hahn}},\ and\ \bibinfo
  {author} {\bibfnamefont {J.}~\bibnamefont {Kublbeck}},\ }\bibfield  {title}
  {\bibinfo {title} {{Feynman rules for fermion number violating
  interactions}},\ }\href {https://doi.org/10.1016/0550-3213(92)90169-C}
  {\bibfield  {journal} {\bibinfo  {journal} {Nucl. Phys. B}\ }\textbf
  {\bibinfo {volume} {387}},\ \bibinfo {pages} {467} (\bibinfo {year}
  {1992})}\BibitemShut {NoStop}%
\bibitem [{\citenamefont {Barela}\ and\ \citenamefont {Monta\~no
  Dom\'\i{}nguez}(2022)}]{Barela:2022sbb}%
  \BibitemOpen
  \bibfield  {author} {\bibinfo {author} {\bibfnamefont {M.~W.}\ \bibnamefont
  {Barela}}\ and\ \bibinfo {author} {\bibfnamefont {J.}~\bibnamefont {Monta\~no
  Dom\'\i{}nguez}},\ }\bibfield  {title} {\bibinfo {title} {{Constraints on
  exotic particle masses from flavor violating charged lepton decays and the
  role of interference}},\ }\href {https://doi.org/10.1103/PhysRevD.106.055013}
  {\bibfield  {journal} {\bibinfo  {journal} {Phys. Rev. D}\ }\textbf {\bibinfo
  {volume} {106}},\ \bibinfo {pages} {055013} (\bibinfo {year} {2022})},\
  \Eprint {https://arxiv.org/abs/2205.08604} {arXiv:2205.08604 [hep-ph]}
  \BibitemShut {NoStop}%
\bibitem [{\citenamefont {Alloul}\ \emph {et~al.}(2014)\citenamefont {Alloul},
  \citenamefont {Christensen}, \citenamefont {Degrande}, \citenamefont {Duhr},\
  and\ \citenamefont {Fuks}}]{Alloul:2013bka}%
  \BibitemOpen
  \bibfield  {author} {\bibinfo {author} {\bibfnamefont {A.}~\bibnamefont
  {Alloul}}, \bibinfo {author} {\bibfnamefont {N.~D.}\ \bibnamefont
  {Christensen}}, \bibinfo {author} {\bibfnamefont {C.}~\bibnamefont
  {Degrande}}, \bibinfo {author} {\bibfnamefont {C.}~\bibnamefont {Duhr}},\
  and\ \bibinfo {author} {\bibfnamefont {B.}~\bibnamefont {Fuks}},\ }\bibfield
  {title} {\bibinfo {title} {{FeynRules 2.0 - A complete toolbox for tree-level
  phenomenology}},\ }\href {https://doi.org/10.1016/j.cpc.2014.04.012}
  {\bibfield  {journal} {\bibinfo  {journal} {Comput. Phys. Commun.}\ }\textbf
  {\bibinfo {volume} {185}},\ \bibinfo {pages} {2250} (\bibinfo {year}
  {2014})},\ \Eprint {https://arxiv.org/abs/1310.1921} {arXiv:1310.1921
  [hep-ph]} \BibitemShut {NoStop}%
\bibitem [{\citenamefont {Hahn}(2001)}]{Hahn:2000kx}%
  \BibitemOpen
  \bibfield  {author} {\bibinfo {author} {\bibfnamefont {T.}~\bibnamefont
  {Hahn}},\ }\bibfield  {title} {\bibinfo {title} {{Generating Feynman diagrams
  and amplitudes with FeynArts 3}},\ }\href
  {https://doi.org/10.1016/S0010-4655(01)00290-9} {\bibfield  {journal}
  {\bibinfo  {journal} {Comput. Phys. Commun.}\ }\textbf {\bibinfo {volume}
  {140}},\ \bibinfo {pages} {418} (\bibinfo {year} {2001})},\ \Eprint
  {https://arxiv.org/abs/hep-ph/0012260} {arXiv:hep-ph/0012260} \BibitemShut
  {NoStop}%
\bibitem [{\citenamefont {Patel}(2017)}]{Patel:2016fam}%
  \BibitemOpen
  \bibfield  {author} {\bibinfo {author} {\bibfnamefont {H.~H.}\ \bibnamefont
  {Patel}},\ }\bibfield  {title} {\bibinfo {title} {{Package-X 2.0: A
  Mathematica package for the analytic calculation of one-loop integrals}},\
  }\href {https://doi.org/10.1016/j.cpc.2017.04.015} {\bibfield  {journal}
  {\bibinfo  {journal} {Comput. Phys. Commun.}\ }\textbf {\bibinfo {volume}
  {218}},\ \bibinfo {pages} {66} (\bibinfo {year} {2017})},\ \Eprint
  {https://arxiv.org/abs/1612.00009} {arXiv:1612.00009 [hep-ph]} \BibitemShut
  {NoStop}%
\bibitem [{\citenamefont {Shtabovenko}(2017)}]{Shtabovenko:2016whf}%
  \BibitemOpen
  \bibfield  {author} {\bibinfo {author} {\bibfnamefont {V.}~\bibnamefont
  {Shtabovenko}},\ }\bibfield  {title} {\bibinfo {title} {{FeynHelpers:
  Connecting FeynCalc to FIRE and Package-X}},\ }\href
  {https://doi.org/10.1016/j.cpc.2017.04.014} {\bibfield  {journal} {\bibinfo
  {journal} {Comput. Phys. Commun.}\ }\textbf {\bibinfo {volume} {218}},\
  \bibinfo {pages} {48} (\bibinfo {year} {2017})},\ \Eprint
  {https://arxiv.org/abs/1611.06793} {arXiv:1611.06793 [physics.comp-ph]}
  \BibitemShut {NoStop}%
\bibitem [{\citenamefont {Fujikawa}\ \emph {et~al.}(1972)\citenamefont
  {Fujikawa}, \citenamefont {Lee},\ and\ \citenamefont
  {Sanda}}]{Fujikawa:1972fe}%
  \BibitemOpen
  \bibfield  {author} {\bibinfo {author} {\bibfnamefont {K.}~\bibnamefont
  {Fujikawa}}, \bibinfo {author} {\bibfnamefont {B.~W.}\ \bibnamefont {Lee}},\
  and\ \bibinfo {author} {\bibfnamefont {A.~I.}\ \bibnamefont {Sanda}},\
  }\bibfield  {title} {\bibinfo {title} {{Generalized Renormalizable Gauge
  Formulation of Spontaneously Broken Gauge Theories}},\ }\href
  {https://doi.org/10.1103/PhysRevD.6.2923} {\bibfield  {journal} {\bibinfo
  {journal} {Phys. Rev. D}\ }\textbf {\bibinfo {volume} {6}},\ \bibinfo {pages}
  {2923} (\bibinfo {year} {1972})}\BibitemShut {NoStop}%
\bibitem [{\citenamefont {Meirose}\ and\ \citenamefont
  {Nepomuceno}(2011)}]{Meirose:2011cs}%
  \BibitemOpen
  \bibfield  {author} {\bibinfo {author} {\bibfnamefont {B.}~\bibnamefont
  {Meirose}}\ and\ \bibinfo {author} {\bibfnamefont {A.~A.}\ \bibnamefont
  {Nepomuceno}},\ }\bibfield  {title} {\bibinfo {title} {{Searching for
  doubly-charged vector bileptons in the Golden Channel at the LHC}},\ }\href
  {https://doi.org/10.1103/PhysRevD.84.055002} {\bibfield  {journal} {\bibinfo
  {journal} {Phys. Rev. D}\ }\textbf {\bibinfo {volume} {84}},\ \bibinfo
  {pages} {055002} (\bibinfo {year} {2011})},\ \Eprint
  {https://arxiv.org/abs/1105.6299} {arXiv:1105.6299 [hep-ph]} \BibitemShut
  {NoStop}%
\bibitem [{\citenamefont {Nepomuceno}\ \emph {et~al.}(2016)\citenamefont
  {Nepomuceno}, \citenamefont {Meirose},\ and\ \citenamefont
  {Eccard}}]{Nepomuceno:2016jyr}%
  \BibitemOpen
  \bibfield  {author} {\bibinfo {author} {\bibfnamefont {A.}~\bibnamefont
  {Nepomuceno}}, \bibinfo {author} {\bibfnamefont {B.}~\bibnamefont
  {Meirose}},\ and\ \bibinfo {author} {\bibfnamefont {F.}~\bibnamefont
  {Eccard}},\ }\bibfield  {title} {\bibinfo {title} {{First results on bilepton
  production based on LHC collision data and predictions for run II}},\ }\href
  {https://doi.org/10.1103/PhysRevD.94.055020} {\bibfield  {journal} {\bibinfo
  {journal} {Phys. Rev. D}\ }\textbf {\bibinfo {volume} {94}},\ \bibinfo
  {pages} {055020} (\bibinfo {year} {2016})},\ \Eprint
  {https://arxiv.org/abs/1604.07471} {arXiv:1604.07471 [hep-ph]} \BibitemShut
  {NoStop}%
\bibitem [{\citenamefont {Corcella}\ \emph {et~al.}(2017)\citenamefont
  {Corcella}, \citenamefont {Coriano}, \citenamefont {Costantini},\ and\
  \citenamefont {Frampton}}]{Corcella:2017dns}%
  \BibitemOpen
  \bibfield  {author} {\bibinfo {author} {\bibfnamefont {G.}~\bibnamefont
  {Corcella}}, \bibinfo {author} {\bibfnamefont {C.}~\bibnamefont {Coriano}},
  \bibinfo {author} {\bibfnamefont {A.}~\bibnamefont {Costantini}},\ and\
  \bibinfo {author} {\bibfnamefont {P.~H.}\ \bibnamefont {Frampton}},\
  }\bibfield  {title} {\bibinfo {title} {{Bilepton Signatures at the LHC}},\
  }\href {https://doi.org/10.1016/j.physletb.2017.09.015} {\bibfield  {journal}
  {\bibinfo  {journal} {Phys. Lett. B}\ }\textbf {\bibinfo {volume} {773}},\
  \bibinfo {pages} {544} (\bibinfo {year} {2017})},\ \Eprint
  {https://arxiv.org/abs/1707.01381} {arXiv:1707.01381 [hep-ph]} \BibitemShut
  {NoStop}%
\bibitem [{\citenamefont {Ramirez~Barreto}\ \emph {et~al.}(2013)\citenamefont
  {Ramirez~Barreto}, \citenamefont {Coutinho},\ and\ \citenamefont
  {Borges}}]{RamirezBarreto:2013edk}%
  \BibitemOpen
  \bibfield  {author} {\bibinfo {author} {\bibfnamefont {E.}~\bibnamefont
  {Ramirez~Barreto}}, \bibinfo {author} {\bibfnamefont {Y.~A.}\ \bibnamefont
  {Coutinho}},\ and\ \bibinfo {author} {\bibfnamefont {J.~S.}\ \bibnamefont
  {Borges}},\ }\bibfield  {title} {\bibinfo {title} {{Vector-bilepton
  Contribution to Four Lepton Production at the LHC}},\ }\href
  {https://doi.org/10.1103/PhysRevD.88.035016} {\bibfield  {journal} {\bibinfo
  {journal} {Phys. Rev. D}\ }\textbf {\bibinfo {volume} {88}},\ \bibinfo
  {pages} {035016} (\bibinfo {year} {2013})},\ \Eprint
  {https://arxiv.org/abs/1307.4683} {arXiv:1307.4683 [hep-ph]} \BibitemShut
  {NoStop}%
\bibitem [{\citenamefont {Barela}\ and\ \citenamefont
  {Pleitez}(2020)}]{PhysRevD.101.015024}%
  \BibitemOpen
  \bibfield  {author} {\bibinfo {author} {\bibfnamefont {M.~W.}\ \bibnamefont
  {Barela}}\ and\ \bibinfo {author} {\bibfnamefont {V.}~\bibnamefont
  {Pleitez}},\ }\bibfield  {title} {\bibinfo {title} {Trimuon production at the
  lhc},\ }\href {https://doi.org/10.1103/PhysRevD.101.015024} {\bibfield
  {journal} {\bibinfo  {journal} {Phys. Rev. D}\ }\textbf {\bibinfo {volume}
  {101}},\ \bibinfo {pages} {015024} (\bibinfo {year} {2020})}\BibitemShut
  {NoStop}%
\end{thebibliography}%


\end{document}